\documentclass[
reprint,
superscriptaddress,
showkeys,
aps,
prd
]{revtex4}

\usepackage[english]{babel}

\usepackage[letterpaper,top=2cm,bottom=2cm,left=3cm,right=3cm,marginparwidth=1.75cm]{geometry}
\usepackage{subfigure}
\usepackage{amsmath}
\usepackage{graphicx}
\usepackage[colorlinks=true, allcolors=blue]{hyperref}
\begin{document}

\title{Anisotropic gravitational waves induced by hypermagnetic fields during the electroweak phase transition epoch}
\author{Mingqiu Li}
\email[]{limingqiu17@mails.ucas.ac.cn}
\affiliation{School of Nuclear Science and Technology, University of Chinese Academy of Sciences, Beijing, P.R.China 100049}
\author{Qi-Shu Yan}
\email[]{yanqishu@ucas.ac.cn}
\affiliation{School of Physics Sciences, University of Chinese Academy of Sciences, Beijing, P.R.China  100049}
\affiliation{Center for Future High Energy Physics, Institute of High Energy Physics, Chinese Academy of Sciences, Beijing 100039, China}
\author{Mei Huang}
\email[]{huangmei@ucas.ac.cn}
\affiliation{School of Nuclear Science and Technology, University of Chinese Academy of Sciences, Beijing, P.R.China  100049}

\begin{abstract}
We study the anisotropies of gravitational waves induced by weak hypermagnetic fields which are randomly distributed and oriented during the electroweak phase transition in the early universe. The theory setup of this study is the standard model plus a real singlet scalar field, which can produce the needed strongly first order electroweak phase transition. Then we investigate how the hypermagnetic fields can convert to magnetic fields and we compute the departure of energy difference between the symmetric phase and the broken phase when the magnetic fields are turned on. It is found that the presence of the hypermagnetic fields can increase the Euclidean action, thus can decrease nucleation temperature, which can lead to a supercool plasma. We point out  that the hypermagnetic field can enhance the gravitational wave production from a first order electroweak phase transition and the inhomogeneity of primordial hypermagnetic field can lead to anisotropies of gravitational waves. By examining three well-motivated distribution of hypermagnetic fields, we calculate the corresponding angular power spectra of stochastic gravitational wave background and find they can be significantly larger than the contributions of the Sachs-Wolfe effects and integrated Sachs-Wolfe effects. Our results show that the anisotropies of gravitational wave could provide a novel probe to the primordial hypermagnetic field in the electroweak phase transition epoch.
\end{abstract}


\maketitle
\section{ Introduction}	
Magnetic fields exist in various scales of the Universe, from planets, stars, and galaxies\cite{2015ASSL407483H,Beck_2015}, to clusters of galaxies\cite{Bonafede2009TheCC} and remote protogalactic clouds. The absence of blazar halos implies magnetic fields with a magnitude $B>10^{-16}$G on Mpc to Gpc scales or $B>10^{-15}$G on kpc scales\cite{Neronov2010EvidenceFS}, while the anisotropy and polarization of the cosmic microwave background imposes an upper bound of $\sim 10^{-9}$G on Mpc scales\cite{Planck:2015zrl,Paoletti:2022gsn}. 

The origin of cosmological magnetic fields remains a mystery. There exist two popular schools of thoughts: on the one hand, their seeds are believed to be created in the early universe (primordial hypothesis), on the other hand, they can be produced during the processes of large-scale structure formation and evolution (astrophysical hypothesis).

The primordial hypothesis is attractive and cosmological magnetic fields can be generated from cosmological phase transitions\cite{articleBianLigong,Zhang:2019vsb}, turbulence\cite{Kulsrud:2007an}, inflation \cite{Martin:2007ue,Ferreira:2013sqa,Campanelli:2013mea}, etc. After their generation, the magnetic fields undergo inverse cascade prior to recombination and then evolve adiabatically. Assuming nowadays the magnitude of cosmological magnetic field is about $10^{-16}-10^{-11}\text{G}$ or so, then it is calculable that at $T\sim 100 \text{GeV}$ epoch the magnetic field strength should be about $10^{18}-10^{23}\text{G}$, or equally $B$ field should be in the range of $(0.14\text{GeV})^2-(44\text{GeV})^2$\cite{cite1111}. Suppose that the primordial magnetic fields are generated by inflation in the very early universe, it is expected that they could play an important role in the successive evolutionary processes, like the electroweak phase transition (EWPT) and quantum chromodynamics (QCD) phase transition. It has been found that QCD matter under external magnetic field shows novel phenomena, including Chiral Magnetic Effect(CME)~\cite{Kharzeev:2007tn,Kharzeev:2007jp,Fukushima:2008xe}, Magnetic Catalysis (MC) in the vacuum ~\cite{Klevansky:1989vi,Klimenko:1990rh,Gusynin:1995nb}, Inverse Magnetic Catalysis (IMC) around the chiral phase transition ~\cite{Bali:2011qj,Bali:2012zg,Bali:2013esa,Chao:2013qpa,Yu:2014sla}, and the magnetic field also dramatically affect hadron properties \cite{Ding:2020jui,Bali:2017ian,Bali:2012jv,Liu:2014uwa,Liu:2015pna,Wang:2017vtn,Mao:2018dqe,Lin:2022ied,Xu:2020yag}.
The gravitation wave and primordial black hole induced by chirality imbalance under magnetic field has been investigated in Ref.\cite{Shao:2022oqw}.

As a working mechanism for baryogenesis\cite{Morrissey:2012db} and a physical process producing stochastic gravitational waves (GWs)\cite{Caprini:2015zlo},  the EWPT has been studied extensively. It is well known that in the Minimal Standard Model (SM) the EWPT is a crossover\cite{Rummukainen:1996sx}, but it can be a strongly first order phase transition when new physics are considered, like the singlet extension of the SM\cite{Vaskonen:2016yiu,Beniwal:2017eik,Beniwal:2018hyi,Alves:2018jsw},  two-Higgs-doublet models\cite{Cline:1996mga,Basler:2016obg,Dorsch:2016nrg},  SUSY models\cite{Huber:2007vva,Huber:2015znp,Demidov:2017lzf}, composite models\cite{Bian:2019kmg,Bruggisser:2018mrt}, seesaw models\cite{Brdar:2018num}, a holographic technicolor model \cite{Chen:2017cyc} and the minimal left-right symmetric model \cite{Li:2020eun} etc. 

Before the electroweak symmetry breaking, the symmetry of the SM is $SU(2)_L\times U(1)_Y$. Instead there is no electromagnetic field but weak gauge field $W$ and hypercharge gauge field $Y$. In the symmetric phase, the non-Abelian $SU(2)_L$ field would be screened over distances larger than the inverse of the magnetic mass, which is nonvanishing and can be expressed as  $m_g \sim g_L^2T$. In contrast, the the hypercharge magnetic field $B_Y$ of the Abelian hypercharge symmetry could be a long-range magnetic force survived in early universe\cite{Elmfors_1998,Kajantie:1996qd,Giovannini:1997eg}. Hypermagnetic field can delay the EWPT\cite{Sanchez:2006tt} or even induce a strongly first order phase transition in the SM if the field strength is sufficient strong\cite{Elmfors_1998,Abedi:2019msi}. An extremely strong hypermagnetic field may even lead to a condensation of gauge field\cite{Chernodub:2022ywg}. Therefore, in this work, we will consider weak hypermagnetic $B_Y\ll m^2_W/e$ background fields (in our numerical calculation, $B_Y$ is no stronger than $(40 \textrm{GeV})^2$ ) which can be generated before the start of the EWPT, and we simply neglect the case with a strong hypermagnetic field $B_Y$ is larger than $(40 \textrm{GeV})^2$. 

During a first order phase transition, vacuum bubbles nucleate, expand, collide, and finally occupy all the space. In the broken phase, both weak charged vector bosons $W$ and neutral vector boson $Z$ obtain their masses and thus are screened in the thermal media, while the electromagnetic component of the original hypermagnetic field remains massless due to the unbroken $U(1)_{em}$ symmetry. Outside the bubbles is the symmetric phase where  hypermagnetic fields $B_Y$ can exist. Inside each a bubble, it is a broken phase and only electromagnetic fields $B$ may survive. It is interesting to investigate how the transition from $B_Y$ to $B$ can affect the EWPT. As one of the key finding in this work, it is found that the hypermagnetic field can decrease the nucleation temperature and produce a supercool plasma. Consequently, during the EWPT, the hypermagnetic can  increase the vacuum energy release, which can lead to a considerable enhancement the GWs emission.

The successful detection of GWs produced by binary black holes \cite{LIGOScientific:2016aoc} demonstrate that GWs could offer a new probe to the early universe, since the GWs can carry the information of phase transitions. Different from  the merge of massive objects which produce a chirp-like GWs in a short period at specific positions, the first order phase transitions produce stochastic gravitational wave background (SGWB) which spread the whole space, fly in all the directions, and last for a long time. 

During a first order phase transition, there are three major physical processes which can produce the GWs: bubble collisions\cite{PhysRevLett.69.2026}, magnetohydrodynamic turbulence\cite{Kosowsky:2001xp}, and sound waves\cite{Hindmarsh:2013xza}. For a wide range of models, during the phase transition, the main dynamics of the GW production is the sound waves, since it may continue to produce GWs long after the bubbles have merged\cite{Hindmarsh:2013xza}. The spectrum of GWs from a first order phase transition has been studied by numerical simulations\cite{Hindmarsh:2013xza} and semi-analytic method\cite{Hindmarsh:2019phv}. 

Similar to the cosmic microwave background, interesting information may be extracted from the anisotropies in SGWB. As far as we know, both the inhomogeneity of GW sources\cite{PhysRevLett.121.201303} and gravitational lensing effects \cite{LISACosmologyWorkingGroup:2022kbp} can lead to anisotropies of the SGWB. In this work, we will study the anisotropies in SGWB caused by the inhomogeneity of GW sources, which are in the form of randomly distributed and oriented external hypermagnegtic fields.

In this paper, we will systematically study the effects of weak hypermagnetic fields to the EWPT. To study the anisotropies in SGWB, we assume inhomogeneous weak hypermagnetic fields as their sources. This assumption can be justified since magnetic fields observed in the Universe have both homogeneous (or uniform) component and inhomogneous (or non-uniform) components \cite{Giovannini:2003yn}. 

It is found in this work that the hypermagnetic field can enhance the GWs emission and the inhomogeneity of hypermagnetic background can induce anisotropies of the SGWB. And it is also found that the magnitudes of GWs can be strong enough to be detected by the future planned space-based GW detectors, like Taiji\cite{Guo:2018npi}, BBO\cite{Corbin:2005ny}, and  DECIGO\cite{Musha:2017usi} for some model parameters. The anisotropies caused by the weak hypermagnetic field can be of the size $10^{-1}$, which is significantly larger than the contributions of Sachs-Wolfe (SW) effects and integrated Sachs-Wolfe (ISW) \cite{LISACosmologyWorkingGroup:2022kbp,Li:2021iva} effects, which is typically of the size $10^{-9}$. 

The rest of the paper is organized as follows. In Section \ref{section_model} we revisit singlet extension of the SM.  How the magnetic field influence EWPT is investigated in Section \ref{section_magnetic}. In Section \ref{sectionGW} the enhancement of SGWB due to magnetic field is study and the anisotropy of SGWB is discussed. Finally, we summarize our result in Section \ref{sectionconclusion}. In Appendix  \ref{magnetic_correction} we discuss the higher order corrections from magnetic field.

\section{A singlet extension of the SM}
\label{section_model}
To achieve a strongly first-order EW phase transition, we consider a singlet extension of the SM. The model includes the SM fields and a singlet real scalar S. By imposing a $Z_2$ symmetry, i.e. $S \to - S$ while the fields of SM is invariant under such a discrete symmetry transformation, this singlet scalar $S$ is only allowed to couple to the SM weak Higgs doublet $H$. The Lagrangian of the SM takes the following simple form \cite{articleBeniwal}
\begin{equation}
	\mathcal{L}_{SM+singlet}=\mathcal{L}_{SM}+\frac{1}{2}\partial_\mu S\partial^\mu S-\frac{\mu_S^2}{2}S^2- \lambda_{HS}|H|^2S^2-\frac{\lambda_S}{4!}S^4,
\end{equation}
where $\mu_S^2$ is a mass parameter, $\lambda_{HS}$ describes the strength of coupling between $S$ and $H$, and $\lambda_S$ is the self-coupling of $S$.  For simplicity, we assume that $\mu_S^2 >0$ and  thus $\langle S\rangle =0$. 

To study the electroweak phase transition, we should calculate the effective potential of Higgs. The vacuum expectation values of Higgs doublet can be put as
\begin{equation}
\label{vev}
\langle H \rangle = \frac{e^{i\theta}}{\sqrt{2}}\left(
\begin{array}{c}
0 \\
v  \\
\end{array}
\right),
\end{equation}
where $\theta$ is a phase factor and $v$ is the vacuum expectation value of the neutral Higgs boson at finite temperature. The effective potential can be cast into the following form \cite{PhysRevD.45.2933,articleCurtin1David,articleBeniwal}:
\begin{equation}
	\begin{aligned}
		\label{eff_P}
V_{eff}(v)&=-\frac{\mu^2}{2}v^2+\frac{\lambda}{4}v^4\\	
&+\sum_{i}\frac{n_i}{64\pi^2}\left[m_i^4(v)\left(\log\frac{m_i^2(v)}{m_{i}^2(v_{EW})}+\frac{3}{2}\right)+2m_i^2(v)m_{i}^2(v_{EW})\right]\\
&+\sum_i \frac{n_iT}{2\pi^2}\int_0^\infty d k k^2 \log\left[1\pm\exp\left(\frac{-\sqrt{k^2+m_i^2(v)}}{T}\right)\right]\\
&-\sum_{i=bosons}\frac{T}{12\pi}\left[(m_i^2(v)+\Pi_i(T))^{3/2}-m_i^3(v)\right]. 
	\end{aligned}
\end{equation}
where the first line is the tree-level potential, the second line is the one-loop corrections at $T=0$, and the third and forth line denote the thermal corrections. The particle index $i$ includes $i=\{W,Z, \chi, h,S,t,\}$ and the corresponding degree of freedom of particles $n_i$ is given as $n_i=\{6,3,3,1,1,-12\}$. We only keep the contribution of top quarks, which have the largest Yukawa coupling to the Higgs bosons, and we have neglected that of light fermions. It is remarkable that only the bosonic contribution can yield to the term $v^3$ in the potential, which is crucial for the first-order phase transition.

The tree-level masses of weak vector boson and scalar Higgs bosons and top quarks as well are given below,
\begin{equation}
	\begin{aligned}
m_W^2&=\frac{g_L^2}{4} v^2, \quad m_Z^2=\frac{g_L^2+g_Y^{ 2}}{4} v^2, \quad m_\chi^2=-\mu^2+\lambda v^2  \\	 
m_h^2 &=-\mu^2+3\lambda v^2\quad	m_S^2=\mu_S^2+\lambda_{HS} v^2 \quad m_t^2=\frac{y_t^2}{2} v^2.
	\end{aligned}
\end{equation}

When we consider thermal corrections, we need thermal masses of these particles which are denoted as $\Pi_i(T)$ in the effective potential given in Eq. (\ref{eff_P}). The thermal masses of vector fields are given as 
\begin{equation}
	\begin{aligned}
		 \Pi_{W^\pm}(T)&=\frac{11}{6} g_L^2 T^2,\\
		\Pi_{W^3,Y}(T) &= \left(
		 \begin{array}{cc}
		 \frac{11}{6} g_L^2 T^2 & 0   \\
		 	0 &  \frac{11}{6} g_Y^2 T^2  \\
		 \end{array}
		 \right),
	\end{aligned}
\end{equation}
which are the same as those of the SM since the scalar field $S$ does not couple to EW gauge fields. And the thermal masses of the two scalars are given as 
\begin{equation}
	\begin{aligned}
		&\Pi_h(T)=\Pi_\chi(T)=T^2\left(\frac{g_Y^{ 2}}{16}+\frac{3 g_L^2}{16}+\frac{\lambda}{2}+\frac{y_t^2}{4}+\frac{\lambda_S}{12}\right), \\
		&\Pi_S(T)=T^2\left(\frac{\lambda_{H S}}{3}+\frac{\lambda_S}{4}\right),
	\end{aligned}
\end{equation}
which is consistent with the relation $\left.\frac{\partial^2V_{eff}}{\partial v^2}\right|_{v=0}=-\mu^2+\Pi_h$.  

\begin{figure}[ht]	
	\centering
	\includegraphics[width=0.7\linewidth]{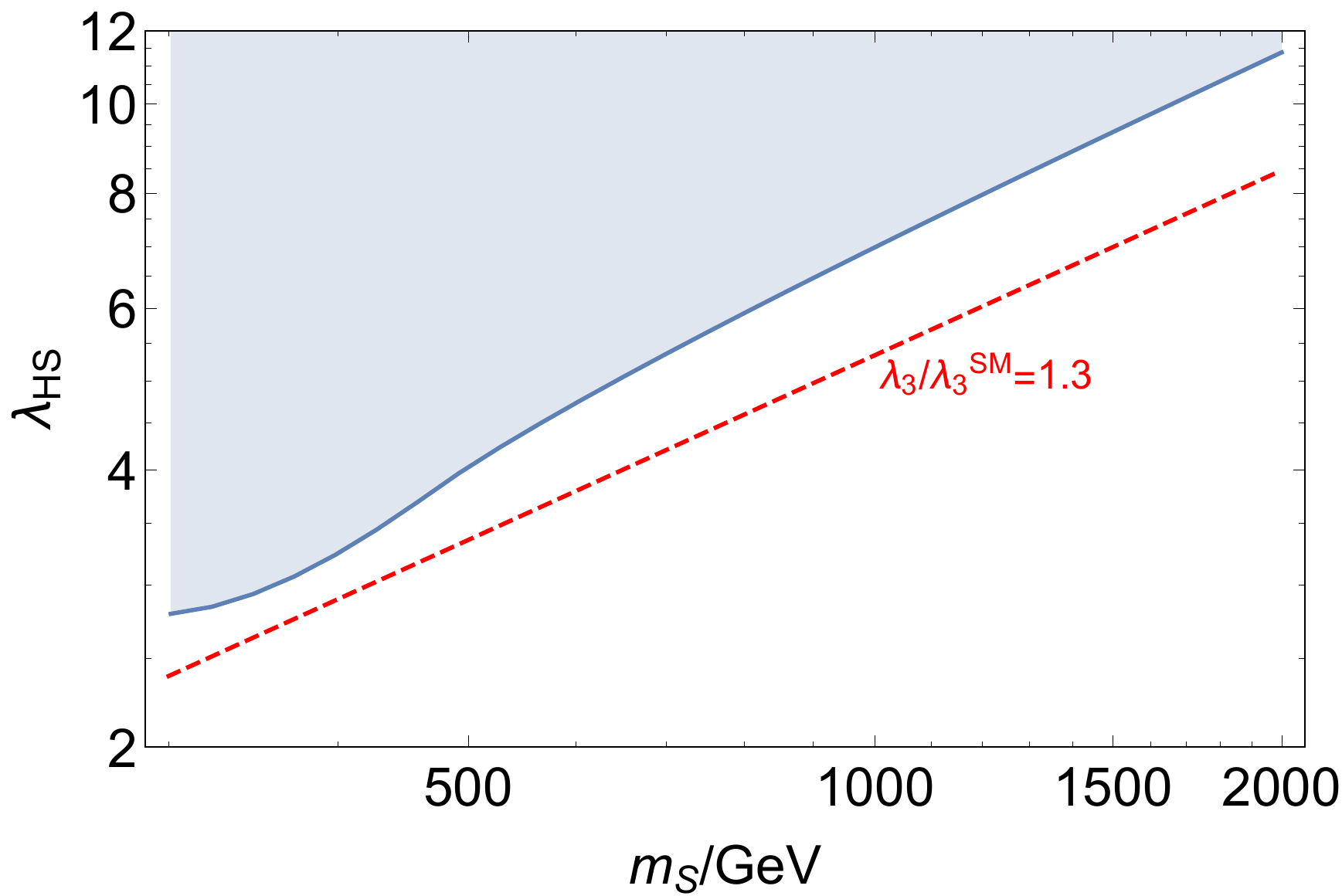}
	\caption{Constraint on the parameter space in the plane $200 \textrm{GeV} < m_S < 2 \textrm{TeV}$ and $2 <\lambda_{HS} < 12$ is shown. The blue shaded area is excluded for $V_{eff}(v_{EW})>V_{eff}(0)$. We have chosen $\lambda_S=1$. The bound that the future LHC can achieved is shown as a dashed line.}
	\label{fig:plotexca}
\end{figure}

When temperature $T=0$, we require that the effective potential should get a minima at $v=v_{EW}\simeq246$GeV, such a condition can constrain the coupling $\lambda_{HS}$ and the mass parameter of $S$, as shown in Figure \ref{fig:plotexca}, where the shaded area is not satisfy such a condition and is excluded.
 
Furthermore, except the SM Higgs, the new scalar $S$ does not couple to other particles of the SM, thus it is difficult to detect it directly by collider data at present. Actually it can serve as a dark matter candidate. When kinematics allows, the SM Higgs boson can decay into a pair of $S$ and can contribute to the invisible Higgs decay. When its mass is above $m_H/2$, its detection becomes challenging.

Nevertheless, its effects can be measured via its radiative corrections to the self-couplings of the SM like Higgs boson. For example, the triple Higgs coupling can be an available probe. At one-loop order, the correction to the triple Higgs coupling of $h^3$ can be modified as
\begin{equation}
	\begin{aligned}
 \lambda_3/\lambda_3^{SM}=\frac{1}{6\lambda_3^{SM}}\left.\frac{\partial^3 V_{eff}(v)}{\partial v^3}\right|_{v=v_{EW},T=0}\simeq 1+0.198\lambda_{HS}^3\left(\frac{100\text{GeV}}{m_{S}}\right)^2	.
	\end{aligned}
\end{equation}
From the di-Higgs boson searches at the LHC \cite{Gouzevitch:2020rwo}, values outside the interval $-1.8<\lambda_3/\lambda_3^{SM}<9.2$ are excluded at 95\% confidence level according to the recent experimental results \cite{Degrassi:2021uik}. The future LHC 14 TeV with an integrated luminosity of 3 $\text{ab}^{-1}$ can provide a more severe constraint $\lambda_3/\lambda_3^{SM}\in (0.7,1.3)$\cite{Barger:2013jfa}, which will be able to detect most of parameter space we interested as Figure \ref{fig:plotexca} shows.

\section{Electroweak phase transition  in an external magnetic field background}
\label{section_magnetic}	
To examine how a homogenous large scale magnetic field can affect the electroweak phase transition, we add a source term for the hyper field tensor to the Lagrangian\cite{Elmfors_1998}(appendix \ref{magnetic_correction}),

\begin{equation}
\label{source}
\mathcal{L}=\mathcal{L}_{SM+singlet}+\frac{1}{2}Y_{\mu\nu}J^{\mu\nu}_{ext}\,,
\end{equation}
where $Y_{\mu\nu}$ is the field strength of the gauge field of $U(1)_Y$ symmetry, which is defined as 

\begin{equation}
Y_{\mu\nu}=\partial_{\mu}Y_{\nu}-\partial_{\nu}Y_{\mu}\,.
\end{equation}

After electroweak symmetry breaking, by using Eq. (\ref{vev}), the Lagrangian of Higgs sector and gauge kinetic terms can be rewritten as 
\begin{equation}
\begin{aligned}
L_{eff}=&\frac{1}{2}\partial_\mu v\partial^\mu v+\frac{v^2}{2}\left(\partial_\mu\theta+\frac{g_L}{2} W^3_\mu-\frac{g_Y}{2} Y_\mu\right)^2\\
&-\frac{1}{4}Y_{\mu\nu}Y^{\mu\nu}+\frac{1}{2}Y_{\mu\nu}J^{\mu\nu}_{ext}-\frac{1}{4}(W^3)_{\mu\nu}(W^3)^{\mu\nu}-V_{eff}(v)	
\end{aligned}
\end{equation}
where we assume vacuum expectation values of charge gauge fields are zero, i.e. $\langle W^1 \rangle=\langle W^2 \rangle=0$, since they are charged under the unbroken $U(1)_{em}$ symmetry.

Then the expectation value of neutral Higgs boson $v$ and the gauge fields $A_\mu$ and $Z_\mu$ should be extreme point of the Lagrangian and satisfy the following conditions
\begin{subequations}
\begin{align}
v^2\partial^\mu\left(\partial_\mu\theta+\frac{g_Z}{2} Z_\mu \right)&=0 \label{eom1}\\
\nabla^2v-\frac{\partial V_{eff}}{\partial v}+v\left(\partial_\mu\theta+\frac{g_Z}{2} Z_\mu \right)^2&=0  \label{eom2}\\	
\partial_{\nu} Z^{\mu \nu}+\sin \theta_{W} \partial_{\nu} J_{\text {ext }}^{\mu \nu}+g_Z v^2\left(\partial^\mu\theta+\frac{g_Z}{2} Z^\mu\right)&=0  \label{eom3}\\
\partial_{\nu} A^{\mu \nu}-\cos \theta_{W} \partial_{\nu} J_{\mathrm{ext}}^{\mu \nu}&=0	\label{eom4}
\end{align}
\end{subequations}
where $g_Z = \sqrt{g_L^2 + g_Y^2}$, $\cos\theta_W = g_L/g_Z$ and $\sin\theta_W=g_Y/g_Z$ and $\theta_W$ is the Weinberg angle. Meanwhile, in Eqs.(\ref{eom1})-(\ref{eom4}), after electroweak symmetry breaking, we redefine the gauge fields from $Y_\mu$ and $W_\mu^3$ to physical vector fields $A_\mu$ and $Z_\mu$ (vice versa) as 
\begin{equation}
\begin{aligned}
\left.\begin{array}{l}
Z_{\mu} \equiv \cos \theta_{W} W_{\mu}^{3}-\sin \theta_{W} Y_{\mu} \\
A_{\mu} \equiv \sin \theta_{W} W_{\mu}^{3}+\cos \theta_{W} Y_{\mu}
\end{array}\right\} \Leftrightarrow\left\{\begin{array}{c}
Y_{\mu}=\cos \theta_{W} A_{\mu}-\sin \theta_{W} Z_{\mu} \\
W_{\mu}^{3}=\sin \theta_{W} A_{\mu}+\cos \theta_{W} Z_{\mu}
\end{array}\right..
\end{aligned}
\end{equation}
For the symmetric phase, we have $v=0$, and the solutions of these equations should respect the following conditions
\begin{equation}
\begin{aligned}
\left\{\begin{array}{l}
Z^{\mu\nu}=-\sin \theta_W J_{ext}^{\mu\nu}  \\
A^{\mu\nu}=\cos \theta_W J_{ext}^{\mu\nu}
\end{array}\right. \Leftrightarrow\left\{\begin{array}{c}
Y^{\mu\nu}=J^{\mu\nu}_{ext} \\
(W^{3})^{\mu\nu}=0
\end{array}\right. .
\end{aligned}
\end{equation}
For the symmetry-broken phase, we have $v>0$, and from Eqs.(\ref{eom1})-(\ref{eom4}) the solutions should satisfy the following relations
\begin{equation}
\begin{aligned}
\left\{\begin{array}{l}
Z^{\mu\nu}=0 \\
A^{\mu\nu}=\cos \theta_W J_{ext}^{\mu\nu}
\end{array}\right. \Leftrightarrow\left\{\begin{array}{c}
Y^{\mu\nu}=\cos^2\theta_w J^{\mu\nu}_{ext} \\
(W^{3})^{\mu\nu}=\cos\theta_w\sin\theta J^{\mu\nu}_{ext}
\end{array}\right. . 
\end{aligned}
\end{equation}

It is obvious that in either phase, the field strength of electromagnetic field $A_{\mu\nu}$ remains constant, while the field strength of $Z$ field  $Z_{\mu\nu}$ depends upon broken or symmetric phase.

Assuming there exists a homogeneous magnetic field background, thus the Gibbs free energy of the system in the broken and symmetric phases can be put as \cite{Elmfors_1998}
\begin{equation}
	\label{free_E}
	\Omega=-L=\left\{\begin{array}{lr}
		V_{eff}(0)-\frac{1}{2}B_{ext}^2 &\text{symmetric phase}\\
		V_{eff}(v)-\frac{1}{2}\cos^2 \theta_w B_{ext}^2&\text{broken phase}
	\end{array}\right . ,
\end{equation}
where $B_{ext}$ denotes the space components of $J_{ext}$ (i.e. $B_{ext}^i = \epsilon^{ijk} J^{jk}_{ext}$ and i,j,k only take space indices and $\epsilon^{ijk}$ is the antisymmetric rank-3 tensor), and the Maxwell
magnetic field is related to it by $B^i = \epsilon^{ijk} A^{jk} =  \cos \theta_W B^i_{ext}$. 
 In the following section, instead of using hypermagnetic fields, we use magnetic fields as our model input, since these two types of fields are simply related. 

The electroweak phase transition of the universe occurs via the nucleation, expansion and merging of EW vacuum bubbles.  The number of bubbles nucleated per time per  volume can be computed and given by the following formula\cite{LINDE1983421,PhysRevD.16.1762}: 
\begin{equation}
	\begin{aligned}
	\Gamma\simeq T^4\exp{\left(-\frac{S_3}{T}\right)}.	
	\end{aligned}
\end{equation}
Here $S_3$ is the Euclidean action of the field configuration which is a saddle point, i.e., solution of Eqs. (\ref{eom1})-(\ref{eom4}). Assuming the background magnetic field is parallel to the z-axis (i.e. $\vec{B}_{ext} = B^3_{ext}$), and using a cylindrical coordinate frame and the unitary gauge (i.e. $\theta=0$), we can put Eq. (\ref{eom2}) and Eq. (\ref{eom3}) into the following forms
\begin{equation}
	\begin{aligned}
		\label{eomzBZ}
		\frac{\partial^2 v}{\partial r^2}+\frac{\partial^2 v}{\partial z^2}	+\frac{1}{r}\frac{\partial v}{\partial r}-\frac{\partial V}{\partial v}-\frac{1}{g_Z^2 \, v^3}(\frac{\partial Z_{12} }{\partial r})^2&=0, \\
		v\frac{\partial^2 Z_{12}}{\partial r^2}+(\frac{v}{r}-2\frac{\partial v}{\partial r})	\frac{\partial Z_{12}}{\partial r}-\frac{g_Z^2}{2}v^3 Z_{12}&=0.
	\end{aligned}
\end{equation}
Meanwhile, Eq. (\ref{eom1}) simply reduces as the covariant Lorentz gauge. To solve these two partial differential equations, we specify the  boundary conditions, which are given as 
\begin{equation}
	\begin{aligned}
		\left.\frac{\partial v}{\partial x_i}\right|_{\vec{x}=0}=0,\quad v(+\infty)&=0,\\
		\left.\frac{\partial Z_{12} }{\partial r}\right|_{r=0}=0,\quad Z_{12} (+\infty)&=-\sin \theta_W  B^3_{ext}.
	\end{aligned}
\end{equation}

In order to evaluate the probability of nucleation of the corresponding bubbles, the solutions of these two differential equations (bounce solutions) can be substituted into the Euclidean action $S_3$, 
\begin{equation}
	S_{3}=\int d x^{3}\left[\frac{1}{2}(\nabla v)^{2}+V_{eff}(v)+\frac{1}{2}Z_{12}^2+\sin\theta_W Z_{12}   B_{ext}^3 +\frac{1}{2}\sin\theta_W^2 (B_{ext}^3)^2\right].
\end{equation}
 The last constant term in the integral is added to guarantee the Lagrangian is zero at symmetric phase.
 
\begin{figure}[htb]
	\centering
	\subfigure{
		\includegraphics[width=0.45\linewidth]{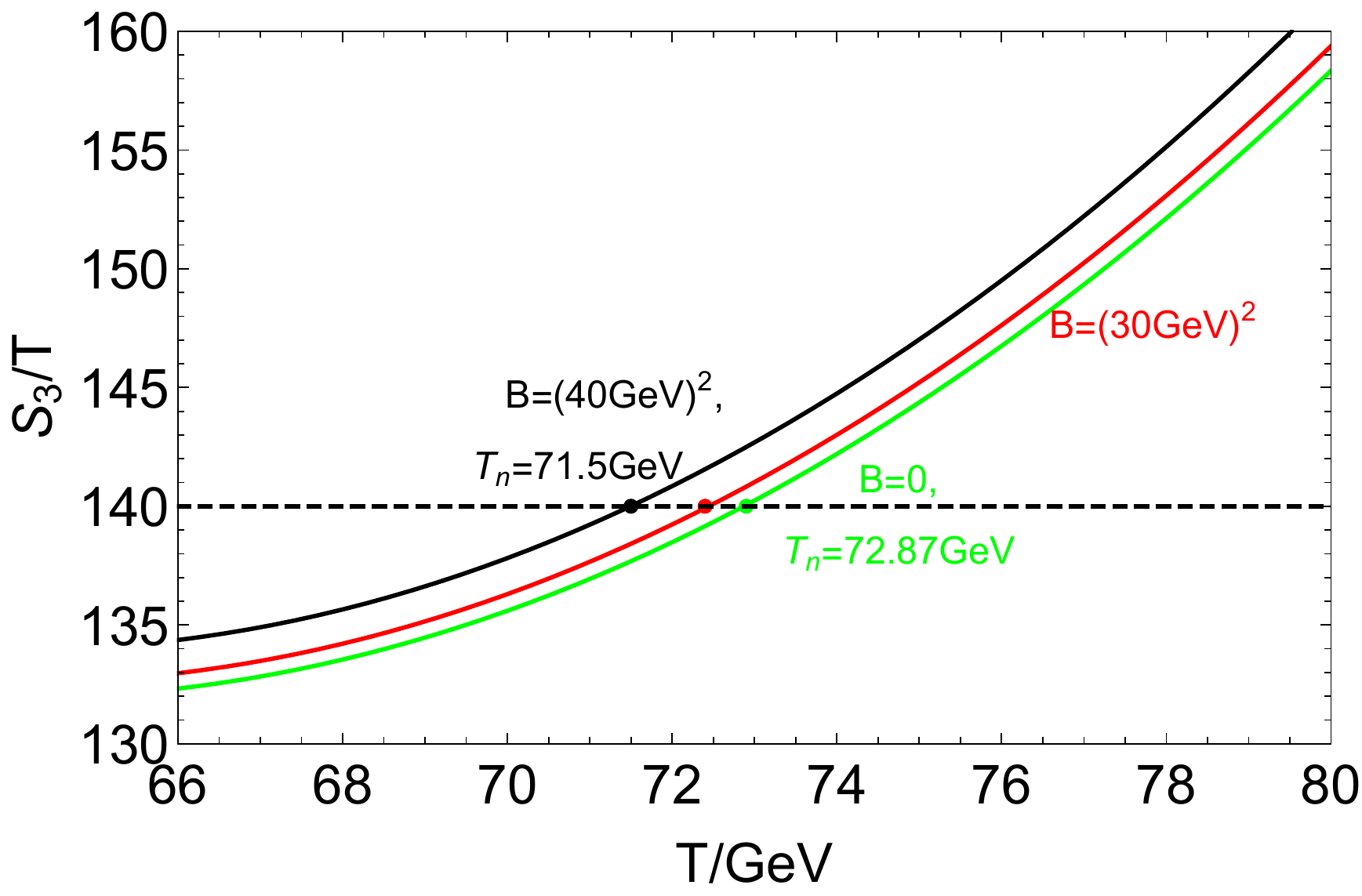} 
	}
	\subfigure{
		\includegraphics[width=0.45\linewidth]{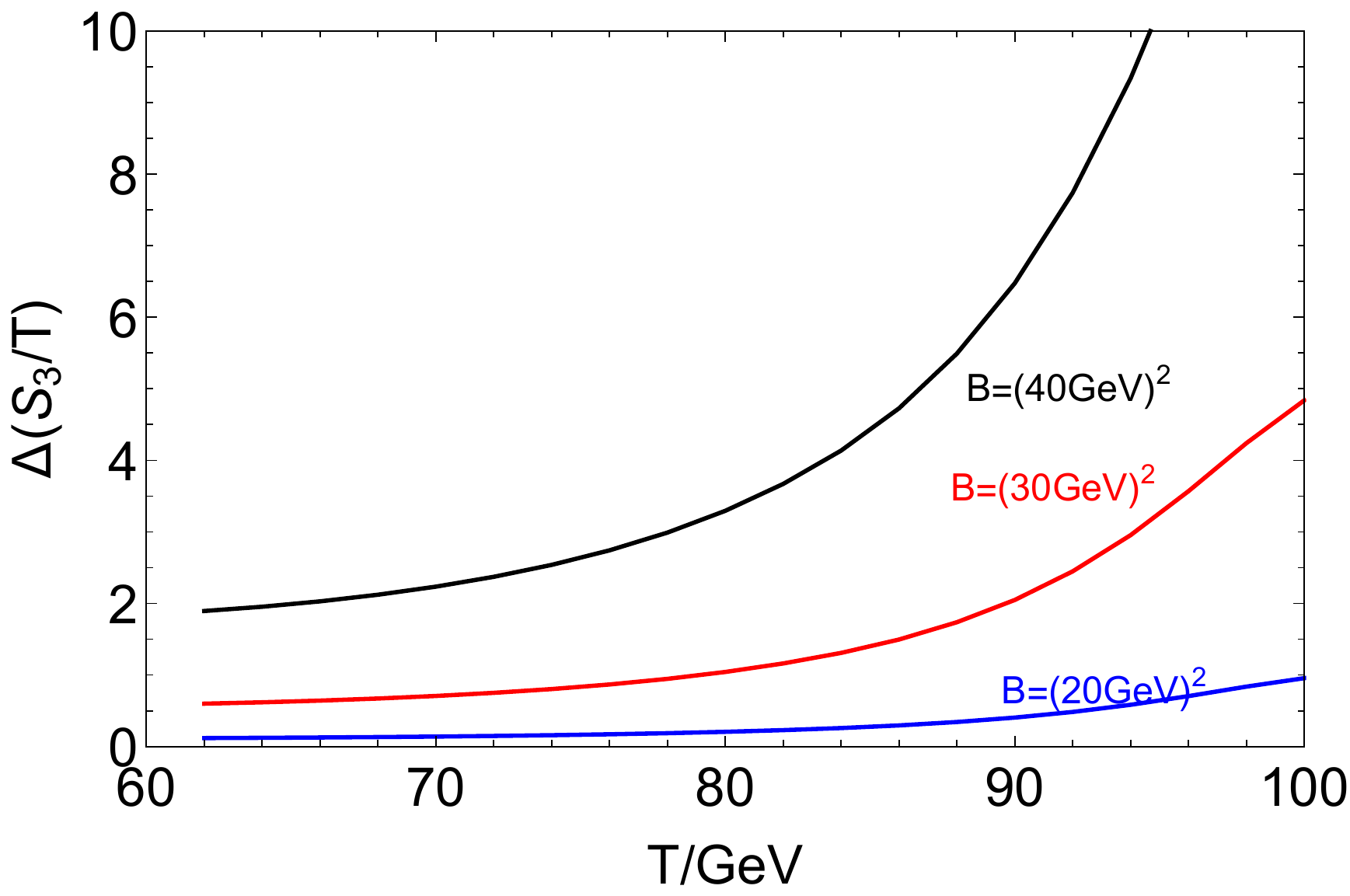}
	}
	\caption{The left plot shows the value of $S_3/T$ at different temperature. The right plot shows the change of $S_3/T$, i.e. $\Delta S_3/T=(S_3/T)_{B\neq 0}-(S_3/T)_{B= 0}$ at different temperature. To show these results, we have fixed other parameters as $\lambda_S=1, \lambda_{HS}=3.67, m_S=500\text{GeV}$.}
	\label{figs3perT}
\end{figure}

The nucleation temperature $T_n$ is defined such that at this temperature there is at least one bubble produced within a Hubble volume. For the electroweak transition, $T_n$ can be well approximated the following equation \cite{Apreda2002GravitationalWF}
\begin{equation}
	\begin{aligned}
		\left.\frac{S_3(T)}{T}\right|_{T=T_n}\simeq 140
	\end{aligned}
\end{equation}
In the left plot of Figure \ref{figs3perT}, we show the values of $S_3/T$  varying of $T$ with $B=(40 \textrm{GeV})^2$, $B=(30 \textrm{GeV})^2$, and $B=0$, respectively. In the right plot, we show the $\Delta S_3/T$ when magnetic field is turned on. We have used $B=\cos \theta_W B_{ext}$ to determine the corresponding Maxwell magnetic field. The $\Delta S_3/T$ is defined as 

\begin{equation}
\left. \frac{\Delta S_3}{T} \right . =\left . \frac{S_3}{T} \right |_{B\neq 0}- \left. \frac{S_3}{T} \right |_{B= 0} \,.
\end{equation}

\begin{figure}[ht]	
	\centering
	\includegraphics[width=0.7\linewidth]{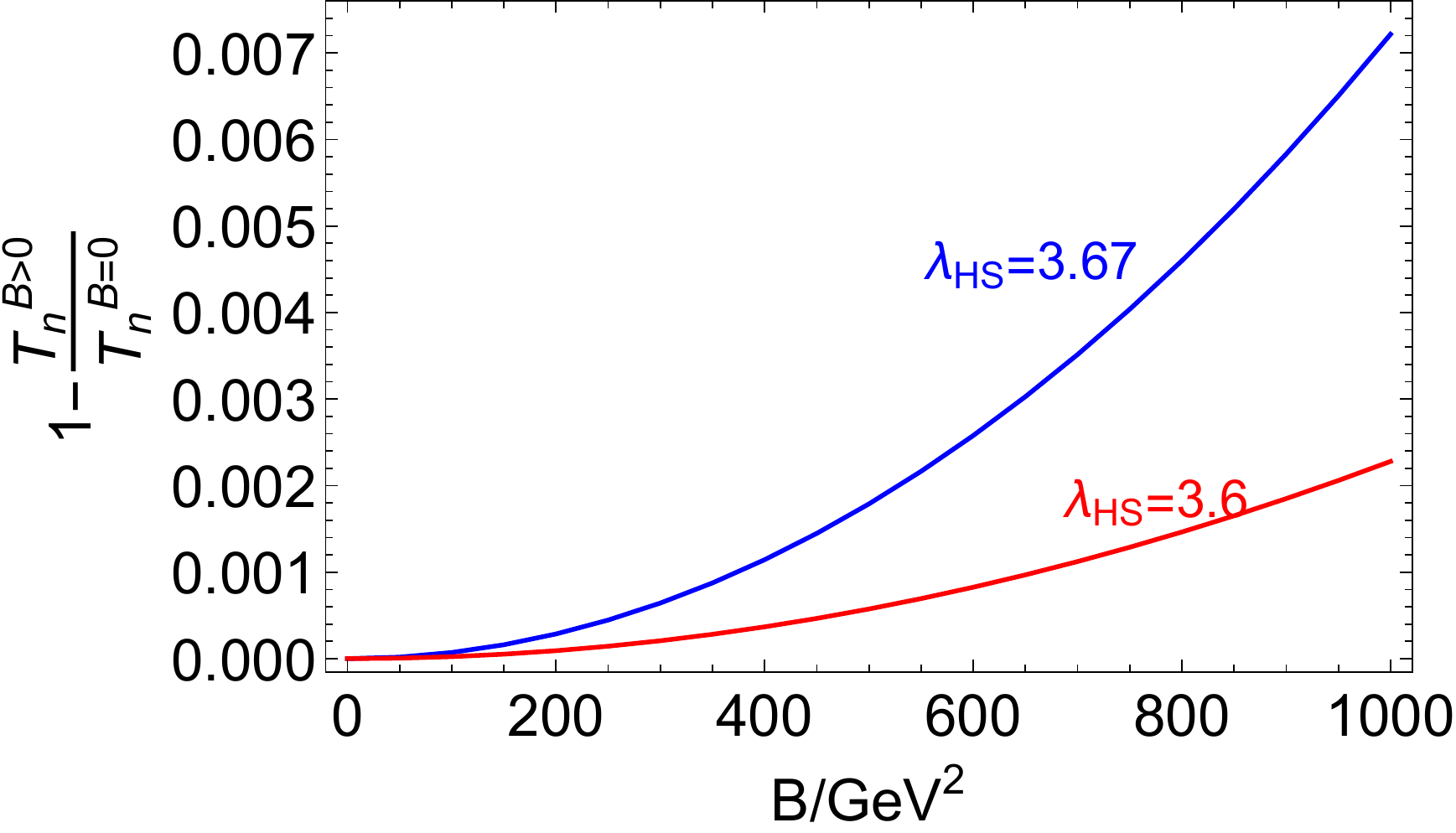}
	\caption{Decrease of nucleation temperature at different background magnetic field. We have fixed other parameters as $\lambda_S=1,  m_S=500\text{GeV}$. When magnetic field is absent, $T_n$ is 95.496GeV for $\lambda_{HS}=3.6$ and $T_n=72.871$GeV for $\lambda_{HS}=3.67$.}
	\label{fig:Tn}
\end{figure}

From the left plot of Figure (\ref{figs3perT}), it is apparent that the background magnetic field can affect the Euclidean action and thus affect the nucleation process. The decrease of nucleation temperature due to the effects of the external magnetic field is shown in Figure (\ref{fig:Tn}). Although the  decrease is less than one percent for $B<1000\text{GeV}^2$, it can have significant effect to the key parameters of phase transition, like $\alpha$ and $\beta$.

 Bounce solution is the critical bubble, of which the surface tension is equal to the pressure difference on the bubble wall. Thus it is the most likely initial configuration. To find the bounce solution when a constant hypermagnetic field exists, in this work we assume that the external hypermagnetic field is weak. In the first step, we use a bounce solution of $v^0$ when external field vanishes as a trial solution, then we substitute the $v^0$ into the second equation of Eq. (\ref{eomzBZ}) to find the corresponding $Z_{12}^0 $. In the second step, we substitute the solution of $Z_{12}^0$ back into the first equation of Eq. (\ref{eomzBZ}) to find the corrected solution of $\delta v$. Then we repeat the first and second steps until the solution converges to specified precision and the procedure stops. In principle, we should use the total $v^t = v^0 + \delta v^1 + \delta v^2 + ...$ and $Z_{12}^0 + Z_{12}^1 + ...$ to compute the $S_3$. In practice, it is found that $\text{max}[\delta v^1] $ ($\text{max}[Z_{12}^1]$) is less than $4 \% \times v^0$ ($Z_{12}^0$)  when the external magnetic field is less than $(40 \textrm{GeV})^2$, therefore, we only use the $v^t = v^0 + \delta v^1$ and $Z_{12}^0$ to compute $S_3$. 
 
\begin{figure}[htb]
	\centering
	\subfigure{
		\includegraphics[width=0.45\linewidth]{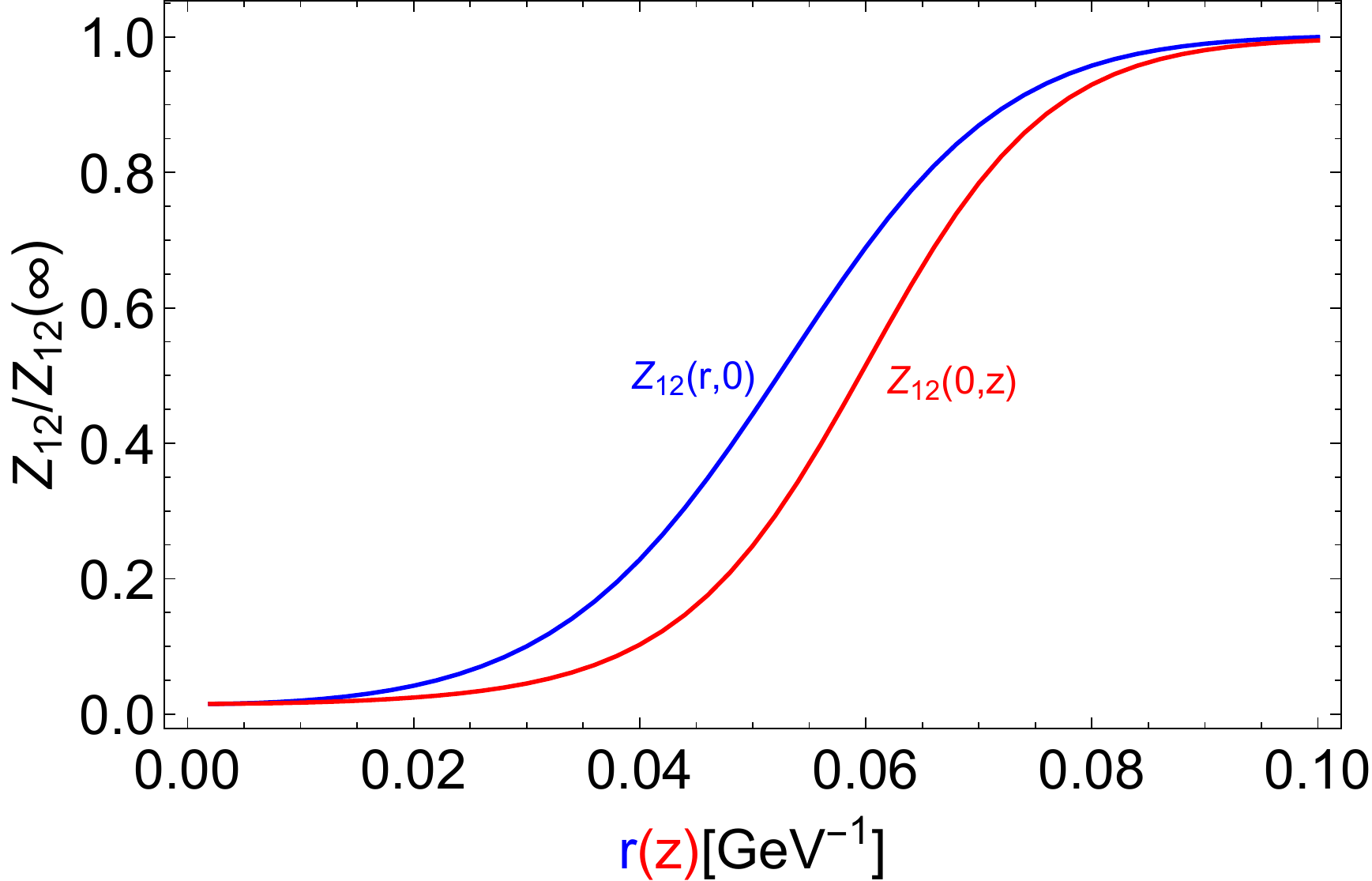} 
	}
	\subfigure{
		\includegraphics[width=0.45\linewidth]{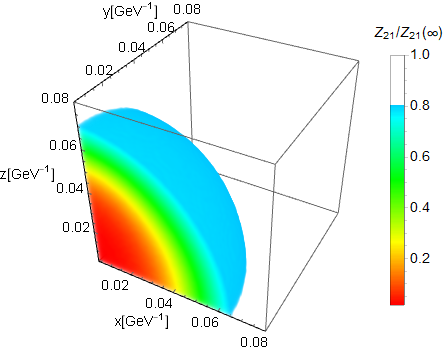}
	}
	\caption{The distribution of $Z_{12}$ around a critical bubble are shown. In the left plot, the values of $Z_{12}$ along $z$ and $r$ directions are shown. In the right plot, 3D values of $Z_{12}$ along the $z$ and $r$ directions are shown where the color bar denotes the values of $Z_{12}$. To show the results, we choose a set of parameters: $\lambda_S=1$, $\lambda_{HS}=3.67$, $m_S=500\text{GeV}$, and $T=72.9\text{GeV}$.}
	\label{fig:plotbubblemagnect}
\end{figure}

 One of the solution of $Z_{12}$ are present as Figure \ref{fig:plotbubblemagnect}. Inside the bubble $Z_{12}$ decrease to zero and outside the bubble $Z_{12}$ approaches to its value in the false vacuum. This is consistent with the fact that in true vacuum $Z$ boson fields obtain mass while in false vacuum $Y$ is massless. After EW bubbles crossing over all space the EWPT ends, and the $Z$ field vanishes everywhere and only magnetic field $A_{ij}$ survives.

\begin{figure}[ht]	
	\centering
	\includegraphics[width=0.7\linewidth]{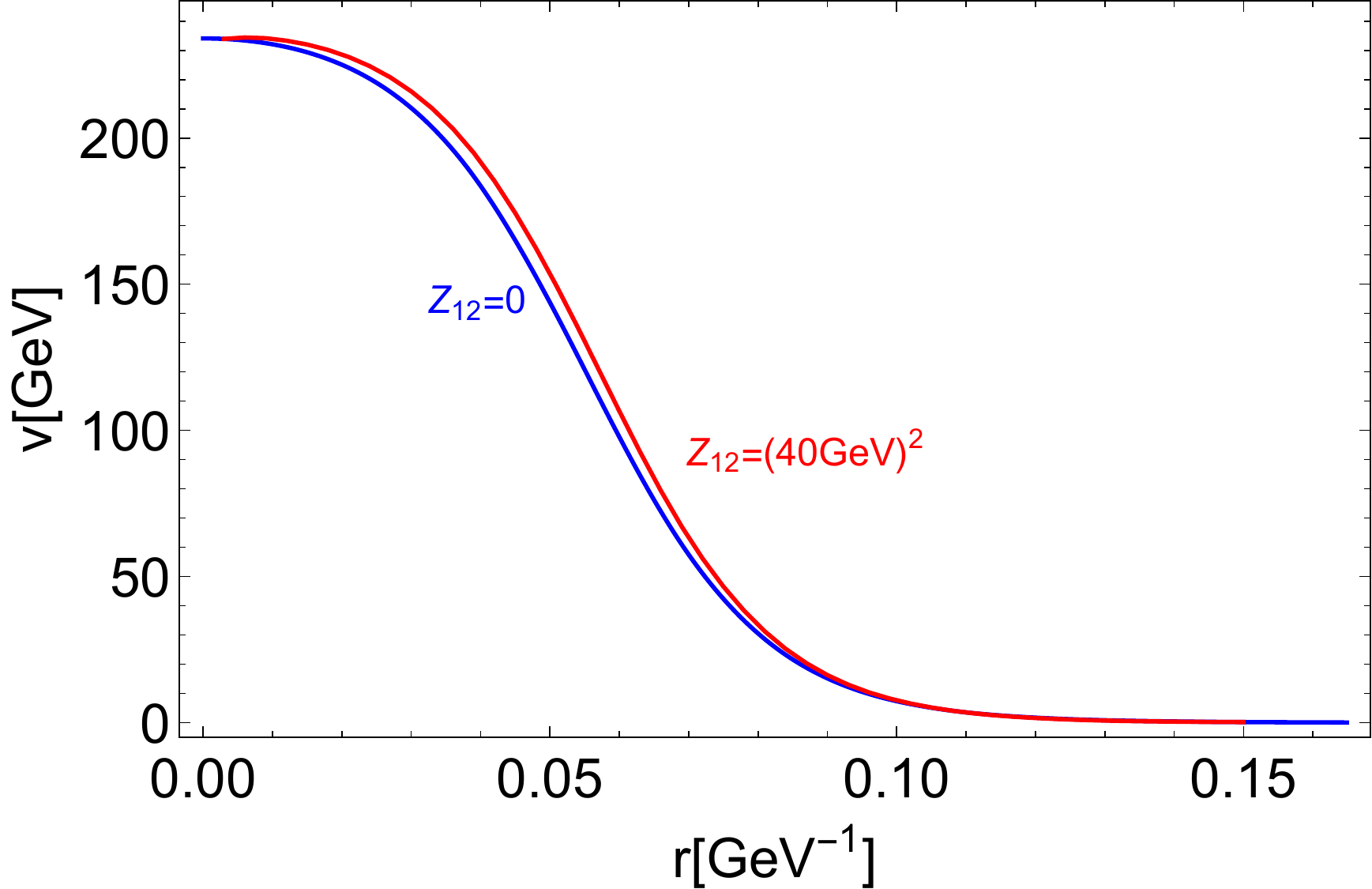}
	\caption{Higgs condensation alone r direction is shown for two cases $Z_{12}=0$ and $Z_{12}=(40 \textrm{GeV})^2$, respectively. Other model  parameters are the same as given in Figure \ref{fig:plotbubblemagnect}.}
	\label{fig:plot6}
\end{figure} 

{In Figure \ref{fig:plot6}, we show the Higgs expectation value $v(r,0)$. From the first equation of Eq. (\ref{eomzBZ}), we can see that the last term can be neglected when $r\rightarrow 0$, thus $v(0,z)$ is not affected by the magnetic field and it should equal to $v(r,0)$ when the magnetic field is zero. As the hypermagnetic field is turned on, to balance the pressure induced by a Z field,  more pressure or less surface tension due to Higgs are needed, and thus $v(r,0)$ should change accordingly. }


\section{Magnitudes and anisotropicity of Gravitational waves}
\label{sectionGW}

According to ref \cite{Hindmarsh2017gnf}, the dominant source of GWs are sound waves created after bubbles collision. The energy density of GWs rescaled by the critical energy of universe is evaluated as \cite{Hindmarsh:2015qta,Guo:2020grp}
\begin{equation}
	\begin{aligned}
			\label{GWsw}
		h^2\Omega_{sw} (f)  \simeq  2.65 \times 10^{-6}\left(\frac{H_n}{\beta}\right)\left(\frac{\kappa_v \alpha}{1+\alpha}\right)^2\left(\frac{100}{g_*}\right)^{1/3} v_w \left( \frac{f}{f_{ sw}}\right) ^3\left[ \frac{7}{4+3 \left(\frac{f}{f_{ sw}}\right)^2} \right]^{7/2} , 
	\end{aligned}
\end{equation}
where the peak frequency $f_{sw}$ is approximated by
\begin{eqnarray}
	f_{sw} & \ \simeq \ & 1.9\times 10^{-2}\frac{1}{v_w}\left(\frac{\beta}{H_n} \right)\left(\frac{T_*}{100\text{GeV}} \right)   \left(\frac{g_*}{100} \right)^{1/6} \text{mHz} ,
\end{eqnarray}
and
\begin{equation}
	\kappa_v\simeq\frac{\alpha}{0.73+0.083\sqrt{\alpha}+\alpha},
\end{equation}
is the fraction of vacuum energy that is converted to bulk motion when bubble wall velocity $v_w\sim 1$ \cite{Espinosa:2010hh,Giese:2020rtr}.
In order to determine magnitudes of the stochastic gravitational waves produced by the EWPT, we should determine two key parameters, $\alpha$ and $\beta$. 

The parameter $\alpha$ describes the fraction of free energy released during the EW phase transition transits to the energy of radiation. As it is known that part of the radiation energy becomes gravitational waves. When the external background magnetic field is taken into account, the parameter $\alpha$ can be put as
\begin{equation}
	\begin{aligned}
	\alpha&=\frac{30}{g_{*} \pi^{2} T^{4} }\left(-\frac{T}{4} \frac{d}{d T} \Delta \Omega+\Delta \Omega\right)
\end{aligned}
\end{equation}
where $\Delta \Omega$ is the free energy released during the EW phase transition and can be expressed
as
\begin{equation}
\label{omegaB}
\Delta \Omega=V_{eff}(0)-V_{eff}(v)-\frac{1}{2}\tan^2\theta_W B^2\,.
\end{equation}

\begin{figure}[ht]	
	\centering
	\includegraphics[width=0.7\linewidth]{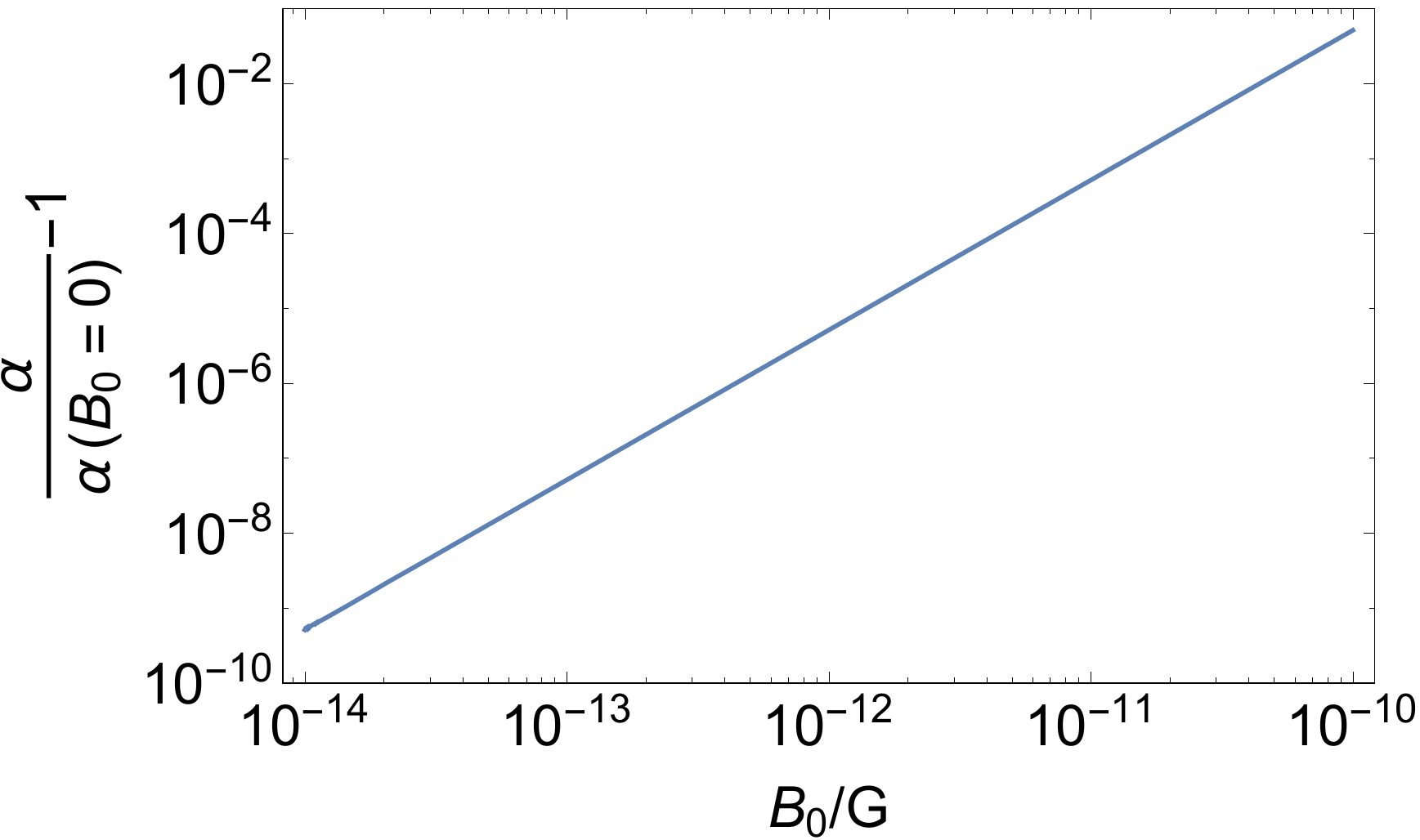}
	\caption{The dependence of $\alpha $ parameter upon magnetic field $B$ is shown. To show the results, we choose a set of parameters: $\lambda_S=1, \lambda_{HS}=3.67, m_S=500\text{GeV}, T=72.9\text{GeV}$.}
	\label{fig:plotdeltaalphab0py}
\end{figure}

\begin{figure}[ht]
	\centering
	\subfigure{
		\includegraphics[width=0.5\linewidth]{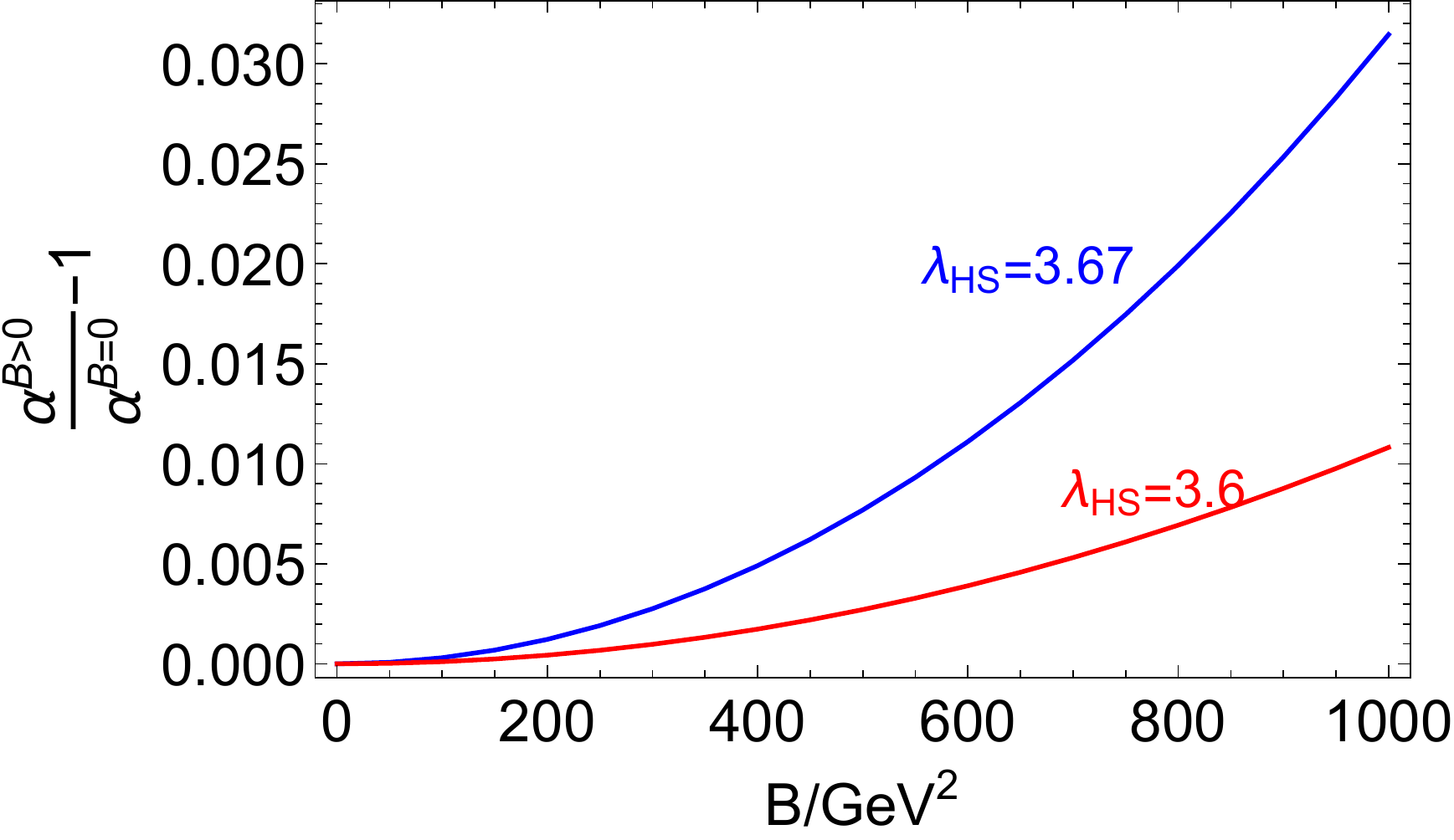} 
	}
	\subfigure{
		\includegraphics[width=0.45\linewidth]{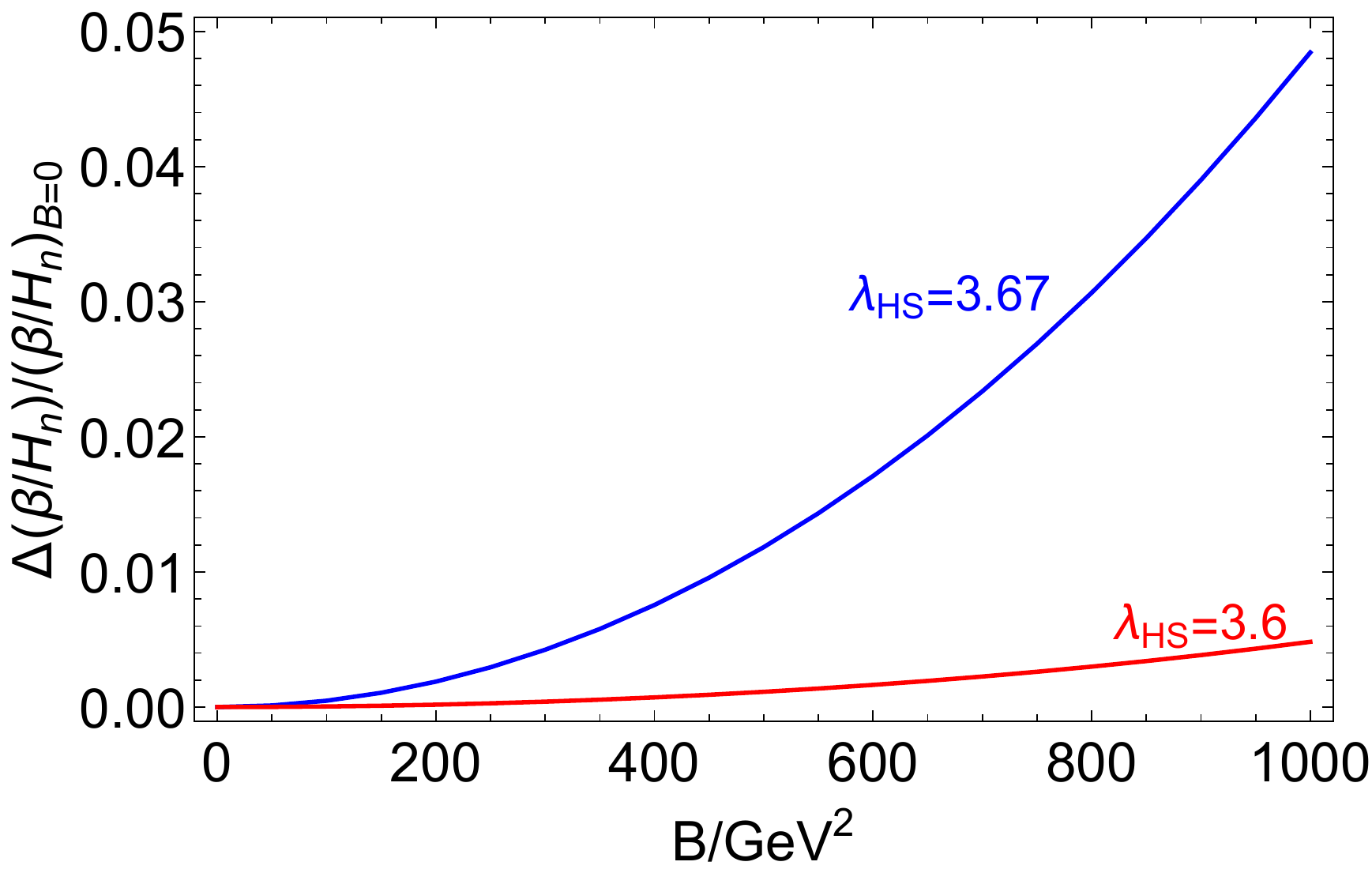}
	}
	\caption{Left plot shows the increase of $\alpha$ at nucleation temperature versus magnetic field. When magnetic field is absent, $\alpha$ is $0.0136$ for $\lambda_{HS}=3.6$ and $\alpha=0.0412$ for $\lambda_{HS}=3.67$. In right plot, $\Delta\beta/H_n:=\left.\beta/H_n\right|_{B=0}-\left.\beta/H_n\right|_{B\neq0}$. When magnetic field is absent, $\beta/H_n$ are 671 or 133 for $\lambda_{HS}=3.6$ and $\lambda_{HS}=3.67$, respectively. Other parameter are the same as figure \ref{fig:Tn}.}
	\label{fig:alphabeta}
\end{figure}

To examine how the last term affect the parameter $\alpha$, how magnetic field evolve as temperature decrease is needed. Prior to the recombination epoch, the magnetic field $B^i = |B |e^i= \epsilon^{ijk}A_{jk}$ ( here $|B|$ denotes the strength of magnetic field and $e^i$ denotes the orientation of the magnetic field) decays in the turbulent plasma. After recombination, the magnetic field evolves adiabatically. Thus for maximally helical magnetic field we can assume that the strength of magnetic field today  $B_0$  is related to the field strength in early universe by \cite{cite1111}
\begin{equation}
	\begin{aligned}
		B(T)=\left(\frac{a}{a_0}\right)^{-2}\left(\frac{\tau}{\tau_{rec}}\right)^{-1/2}B_0.
		\label{BTB0}
	\end{aligned}
\end{equation}
Here $a$ is the  scale factor and $\tau$ is the conformal time. for $B_0$ in $10^{-15}$G to $10^{-11}$G, at nucleation temperature $T_n\simeq 72.9$GeV the corresponding magnetic field $B(T)$ is about $(0.31\text{GeV})^2\sim (31\text{GeV})^2$. Using Eq. (\ref{BTB0}), how the last term in Eq. (\ref{omegaB}) affect  $\alpha$ is  showed in Figure \ref{fig:plotdeltaalphab0py}. In Figure \ref{fig:alphabeta} we show the increase of $\alpha$ at nucleation temperature. Comparing  to the $\tan^2\theta_W B^2$ term in Eq. (\ref{omegaB}), the decrease of nucleation temperature has more significant effect.

The inverse duration of phase transition $\beta$ is a crucial parameter to describe the rate of  phase transition, which is given by
\begin{equation}
	\begin{aligned}
		\frac{\beta}{H_n}=T\left.\frac{d (S_3/T)}{d T}\right|_{T=T_n}
	\end{aligned}
\end{equation}
where $H_n$ is Hubble parameter at the nucleation temperature $T_n$. It is found that the external field can affect the value of $\beta/H_n$, As the right plot in Figure \ref{fig:alphabeta} shows. 

\begin{figure}[ht]
	\centering
	\subfigure{
		\includegraphics[width=0.4\linewidth]{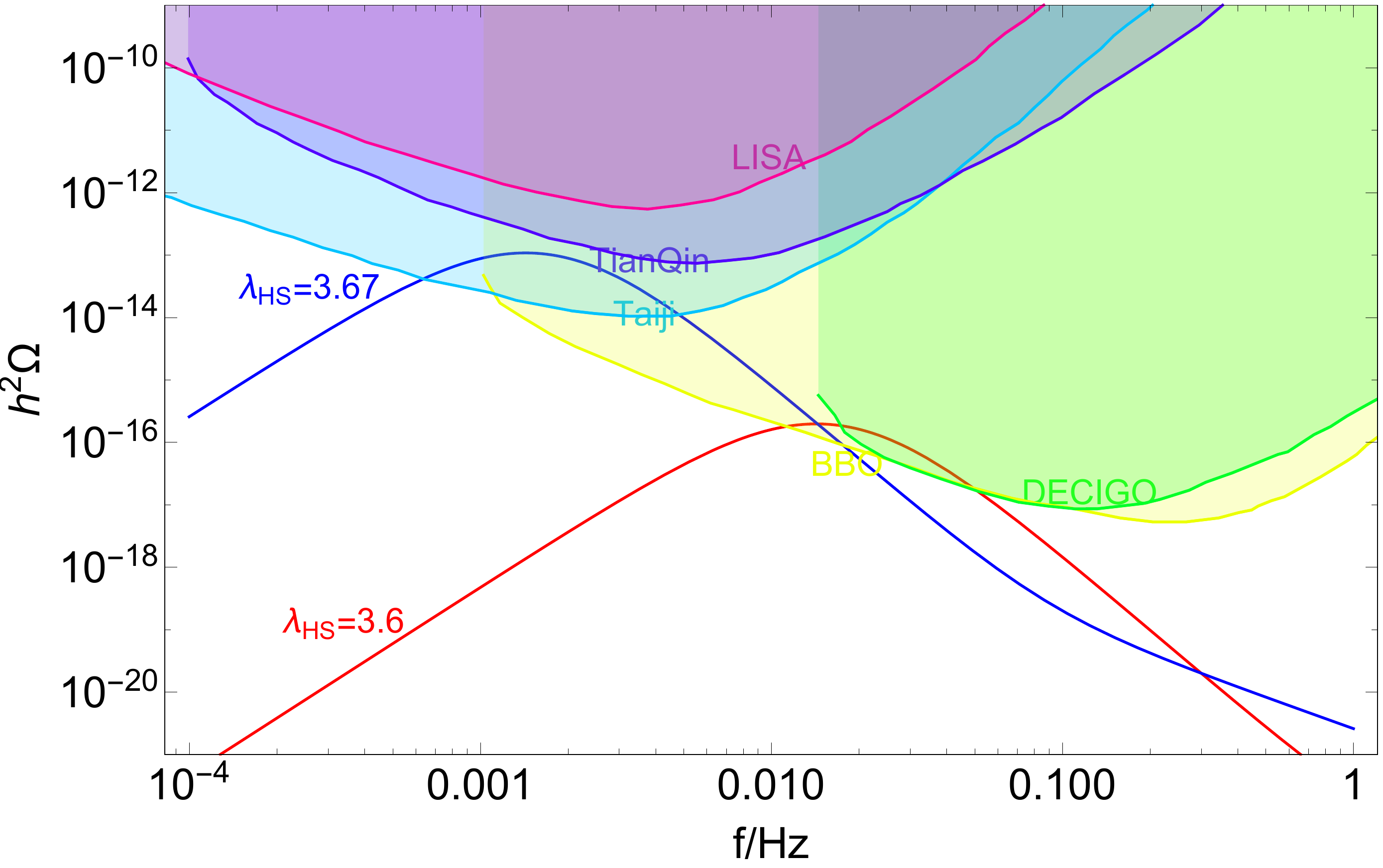} 
	}
	\subfigure{
		\includegraphics[width=0.45\linewidth]{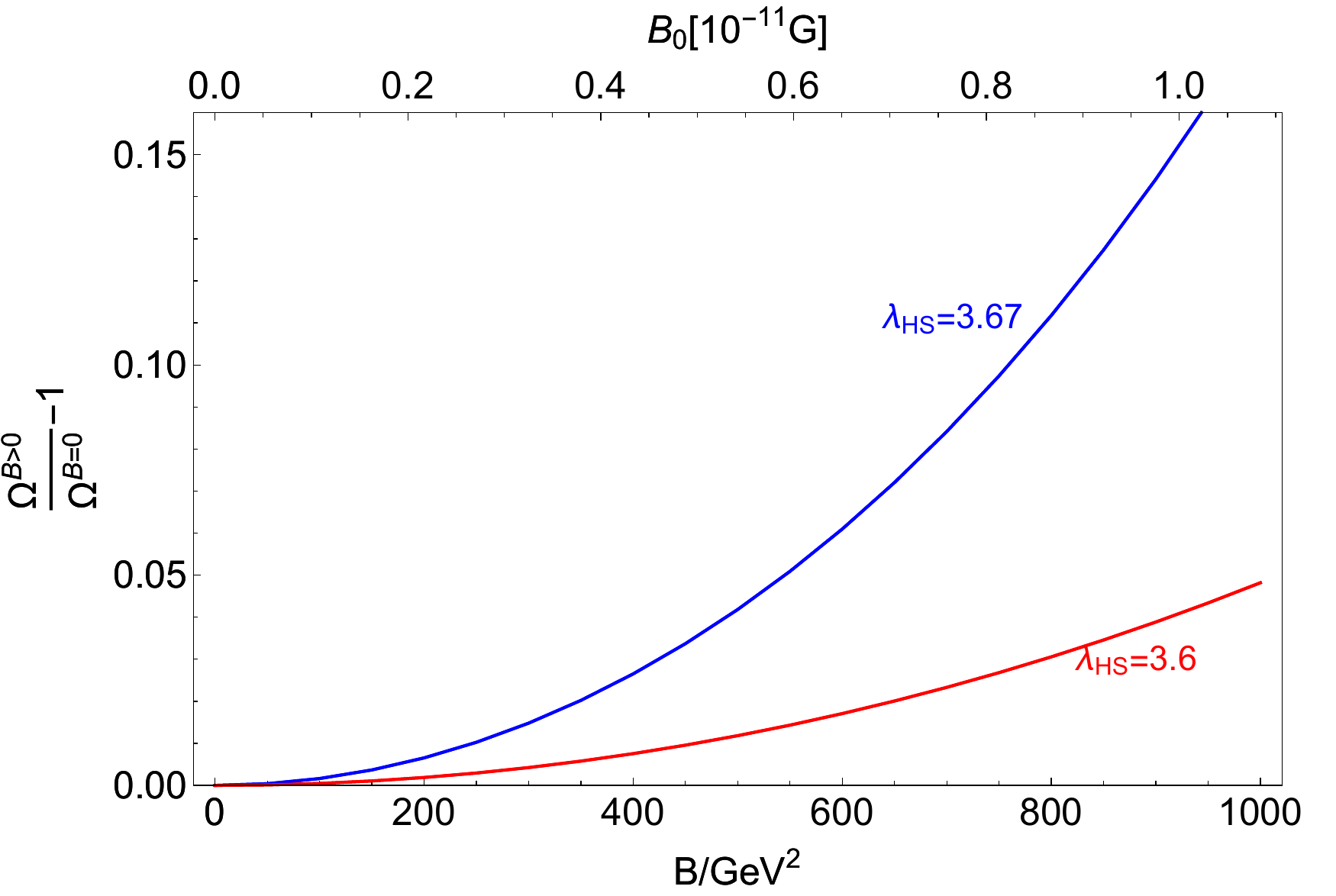}
	}
	\caption{Left plot is to show the energy densities of GWs with $B=0$ with two cases $\lambda_{HS}=3.67$ and $\lambda_{HS}=3.6$. Right plot is to show the effect of magnetic field. To show the results, we choose a set of parameters: $\lambda_S=1$ and $m_S=500\text{GeV}$. }
	\label{wolf2}
\end{figure}

In Figures \ref{wolf2}, we show the spectra of GWs when $B=0$ and the effects of external magnetic fields. It is observed that the magnitude is quite dependent upon the model parameter $\lambda_{HS}$, the maximum magnitude with $\lambda_{HS}=3.67$ can be two orders larger than that of $\lambda_{HS}=3.60$, as shown in the left plot. From the right plot, it is clear that when the model parameter $\lambda_{HS}$ is fixed, the larger the strength of external magnetic field, the larger is the  magnitude of energy density of GWs. For example, when $\lambda_{HS}=3.67$ and $B=800/\textrm{GeV}^2$, the magnitude can be enhanced by $10\%$. Such an enhancement can be attributed to the increase of $\alpha$ and $\beta/H_n$ shown in Figure \ref{fig:alphabeta}.

In the numerical analysis above, we have assumed a homogeneous magnetic field background. But in the cosmological scale,  the magnetic fields are inhomogeneous \cite{Vachaspati:2020blt}. Our study is valid only if $\lambda_B \gg R_b$, where $\lambda_B$ denotes the peak length of magnetic field spectrum and $R_b$ is the average distance between bubbles which can be estimated as $(8\pi)^{1/3}v_{w}/\beta$. Below we investigate the consequence of the inhomogeneous magnetic fields to the GWs. For the sake of  convenience, we will fix the model parameter $\lambda_{HS}=3.67$ and $m_S=500 \textrm{GeV}$.

As it is well-known, the inhomogeneity of hypermagnetic fields necessarily leads to the inhomogeneity of quantity $S_3$. If we assume that inside the plasma everywhere has the same termperature $T$, then a place A with a stronger hypermagnetic field can have a larger $S_3(A)$. In contrast, a place B with a weaker hypermagnetic field can have a smaller $S_3(B)$, i.e. $S_3(B)/T < S_3(A)/T$. Then the EWPT at the place B can occur earlier than it does at the place A, thus the anisotropies of GWs can be generated. Similar to its counterpart in the cosmic microwave background,  anisotropies of GWs can be decomposed in spherical harmonics:
\begin{equation}
	\delta_{\mathrm{GW}}\left(\hat{n}\right)=\frac{\Omega(\hat{n})-\bar{\Omega}}{\bar{\Omega}}=\sum_{\ell} \sum_{m=-\ell}^{\ell} \delta_{\mathrm{GW}, \ell \mathrm{m}} Y_{\ell m}(\hat{n}).
\end{equation}	
Under the assumption of statistical isotropy, the multipole coefficients can be defined as
\begin{equation}
	\left\langle\delta_{\mathrm{GW}, \ell \mathrm{m}} \delta_{\mathrm{GW}, \ell^{\prime} \mathrm{m}^{\prime}}^{*}\right\rangle=C_{\ell}  \delta_{\ell \ell^{\prime}} \delta_{ m m^{\prime}}.
\end{equation}
In their nature, the anisotropies of GWs come from  perturbations of spacetime and the inhomogeneous emission of GWs\cite{LISACosmologyWorkingGroup:2022kbp,Bartolo:2019yeu}. The latter contribution to $C_l$ is
\begin{equation}
	C_{\ell}=4 \pi \int \frac{d k}{k}\left[j_{\ell}\left(k\left(\eta_{0}-\eta_{\mathrm{in}}\right)\right)\right]^{2} P_{\delta }( k)\,,
\end{equation}	
where $j_l$ is spherical Bessel function. $\eta_{in}$ is the  time when GWs produced and $\eta_0$ is the age of universe.
  The initial power spectrum of $\delta_{GW}$ is defined by
\begin{equation}
\label{Psdelta}
	\left\langle\delta_{GW}\left(\eta_{\text {in }}, \vec{k}\right) \delta_{GW}^{*}\left(\eta_{\text {in }}, \vec{k}^{\prime}\right)\right\rangle=\frac{2 \pi^{2}}{k^{3}} P_{\delta }( k)(2 \pi)^{3} \delta\left(\vec{k}-\vec{k}^{\prime}\right)\,,
\end{equation}
and $\delta_{GW}(\vec{k})$ is the Fourier transformation of $\delta_{GW}(\vec{x})$.  
The  spectrum $P_\delta$ is depended on magnetic fields.
Cosmological magnetic fields are often characterized by  power spectrum and helicity spectrum, which are defined by the correlator:
\begin{equation}
	\label{correlatorB}	
	\langle B^{i}_{\bf{k}}B^{*j}_{{\bf{k}}^{\prime}}\rangle=\left[{\frac{E_{M}(k)}{4\pi k^{2}}}p_{i j}+i\epsilon_{i j l}k^{l}{\frac{H_{M}(k)}{8\pi k^{2}}}\right](2\pi)^{6}\delta({\vec{k}}-{\vec{k}}^{\prime})\,,
\end{equation}
where $p_{i j}=\delta_{i j}-\hat{k}_{i}\hat{k}_{j}.$ $E_M(k)$  describe the energy density $\rho_B$ of magnetic fields in k space and $H_M(k)$ is related to the helicity $h$. The quantities $\rho_B$ and $h$ are  
\begin{equation}
	\label{rhohB}
	\rho_{B}=\int d k\,E_{M}(k),\quad  h=\int d k\,H_{M}(k)	.
\end{equation}
The two spectra obey the relationship $E_{M}(k)\geq k|H_{M}(k)|/2$ and magnetic fields  are called maximally helical fields if the equal sign is achieved.
For weak magnetic fields, we can model the anisotropies of GWs as $\delta_{GW}(x)= CB^2(x)$. As shown in Figure \ref{wolf2}, for the red line we take $C=4.06\times 10^{-8}\text{GeV}^{-4}$, and for the blue line we take $C=1.61\times10^{-7}\text{GeV}^{-4}$. In Fourier space, the perturbation of energy caused by magnetic fields can be expressed as 
\begin{equation}
	\begin{aligned}
		\delta_{GW}(\vec{k})&=C\int d^3x	B^2(x)e^{i\vec{k}\cdot\vec{x}}\\
		&=C\int{d^3x \frac{d^{3}k_1}{(2\pi)^{3}}}\frac{d^{3}k_2}{(2\pi)^{3}}B^i_{\bf{k}_1}B^i_{\bf{k}_2}e^{-i\left(\bf{k}_1+{k}_2-{k}\right)\cdot\bf{x}}\\
		&=C\int{ \frac{d^{3}k_1}{(2\pi)^{3}}}B^i_{\bf{k}_1}B^i_{\bf{k}-{k}_1}\,.\\
	\end{aligned}
\end{equation}
Thus, the correlator of $\delta_{GW}$ can be put as 
\begin{equation}
	\begin{aligned}
		\langle \delta_{GW}(\vec{k})\delta_{GW}^*(\vec{k^{'}} )\rangle &=C^2\int \frac{d^{3}k_1}{(2\pi)^{3}} \frac{d^{3}k_2}{(2\pi)^{3}} \langle B^i_{\bf{k}_1}B^i_{\bf{k}-{k}_1}B^{*j}_{\bf{k}_2}B^{*j}_{\bf{k^{'}}-{k}_2}\rangle\\ 
		&=C^2	\int \frac{d^{3}k_1}{(2\pi)^{3}} \frac{d^{3}k_2}{(2\pi)^{3}} \left(\langle B^i_{\bf{k}_1}B^i_{\bf{k}-{k}_1}\rangle\langle B^{*j}_{\bf{k}_2}B^{*j}_{\bf{k^{'}}-{k}_2}\rangle\right.\\
		&+\langle B^i_{\bf{k}_1}B^{*j}_{\bf{k}_2}\rangle\langle B^i_{\bf{k}-{k}_1}B^{*j}_{\bf{k^{'}}-{k}_2}\rangle
		+\left.\langle B^i_{\bf{k}_1}B^{*j}_{\bf{k^{'}}-{k}_2}\rangle\langle B^i_{\bf{k}-{k}_1}B^{*j}_{\bf{k}_2}\rangle\right)\\
		&=C^2(2\pi)^4\delta({\vec{k}}-{\vec{k}}^{\prime})\\
		&\times\int d^3k_1\left[\frac{E_M(k)}{k^2}\frac{E_M\left(\left|\vec{k}-\vec{k}_1\right|\right)}{\left|\vec{k}-\vec{k}_1\right|^2}-\frac{H_M(k)}{2}\frac{H_M\left(\left|\vec{k}-\vec{k}_1\right|\right)}{2\left|\vec{k}-\vec{k}_1\right|^2}\right]\\
		&=2C^2(2\pi)^5\delta({\vec{k}}-{\vec{k}}^{\prime})\left[\frac{E_M(k)}{k^2}\rho_B-\frac{H_M(k)}{4}h\right]\,.
	\end{aligned}
\end{equation}
We have used Eq. (\ref{correlatorB}) and Eq. (\ref{rhohB}) in the last two steps.
Comparing with the definition of $P_\delta$ given in Eq. (\ref{Psdelta}), we arrive at the following relation
\begin{equation}
	\begin{aligned}
		P_\delta(k)=4C^2kE_M(k)\rho_B-C^2k^3H_M(k)h.
	\end{aligned}
\end{equation}
As shown in Figure \ref{wolf2}, when $\lambda_{HS}=3.67$, $\delta_{\mathrm{GW}}(\hat{n}):=\delta_{GW}(\hat{x}(\eta_0-\eta_{in}))$ can change in the range $[0,  0.15]$. Approximately, the power spectra of GWs $P_{\delta}$ can be estimated as $\delta^2_{GW}(\vec{x})$. The detail of $P_{\delta}$ is dependent upon the spectra of hypermagnetic field. Below, we examine the anisotropies of GWs by considering the following three power spectra for magnetic fields:
\begin{itemize}
 \item Scenario 1:  $P_\delta=A^2$ 
\item Scenario 2:  $P_\delta(k)=A^2\exp[-(k\lambda_B -2\pi)^2]$ 
 \item Scenario 3:  $P_\delta(k)=4C^2kE_M(k)\rho_B,$



where $E_M(k)=\left\{
\begin{array}{cc} 	
		\frac{10 \rho _B }{17 k_B}\left(\frac{k}{k_B}\right)^4 & k<k_B \\
		\frac{10 \rho _B }{17 k_B}\left(\frac{k}{k_B}\right)^{-5/3} & k>k_B \\
\end{array}\right. \,,\, \rho_BC=\frac{17}{20}A$.
\end{itemize}
\begin{figure}[ht]	
	\centering
	\includegraphics[width=0.7\linewidth]{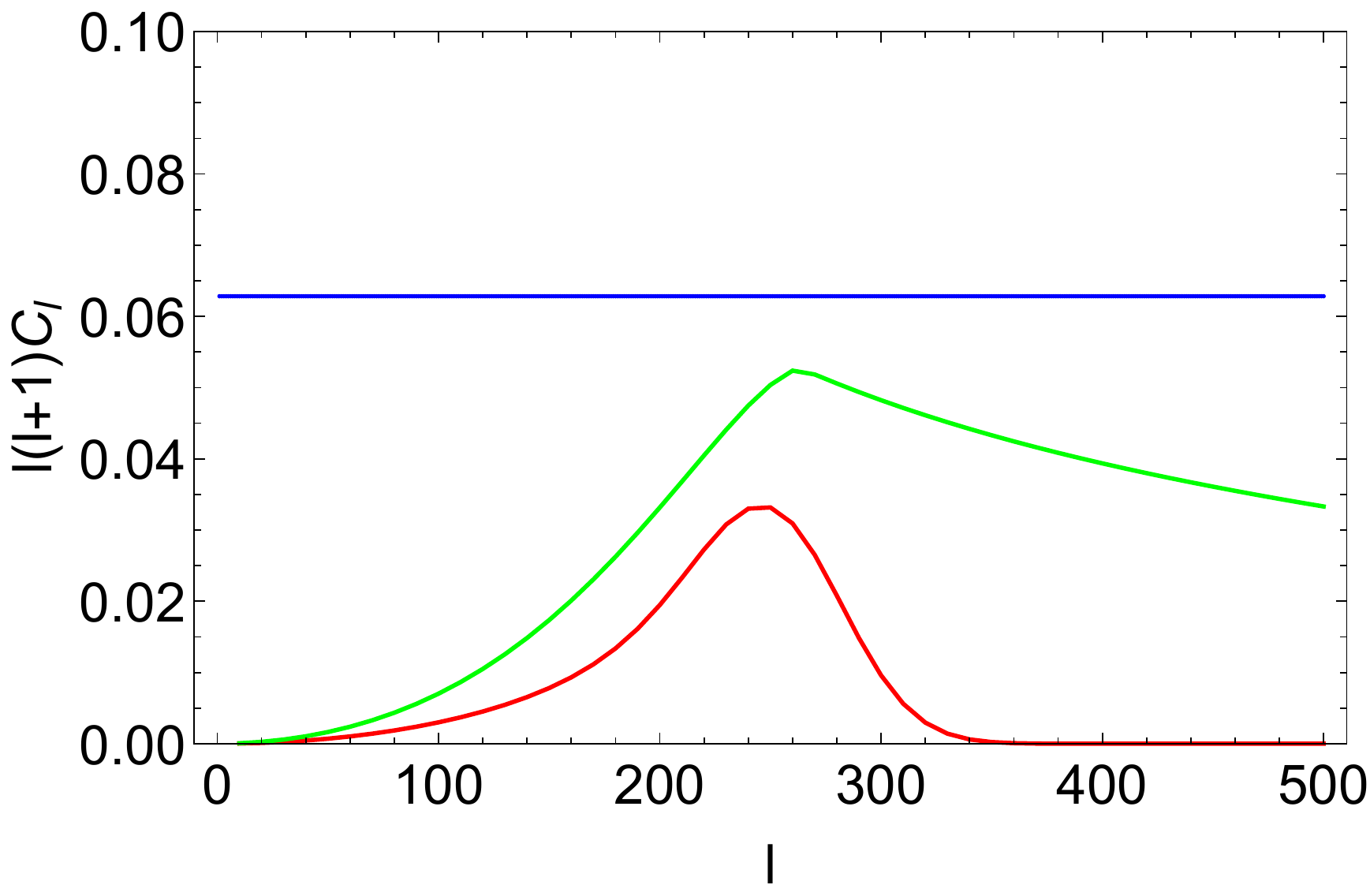}
	\caption{The angular power spectra of the stochastic gravitational
wave background are shown. The blue, red and green lines correspond to  scenario 1, scenario 2, and scenario 3, respectively. To make the plot, we choose the following parameters $A=0.1, \lambda_B=2\pi/k_B=100\text{Mpc}$.}
	\label{fig:plotcl}
\end{figure} 

For all three scenarios, we choose the parameters $A=0.1, \lambda_B=2\pi/k_B=100\text{Mpc}$.
In Figure \ref{fig:plotcl}, we show the angular power spectra of GWs in these scenarios.

The first scenario corresponds a constant spectra of magnetic fields. If magnetic fields were produced on large coherence scales during inflation\cite{Campanelli:2013mea,Ferreira:2013sqa,Martin:2007ue}, the spectra of magnetic fields can be scale invariant, which can lead to a constant power spectra $P_\delta$ in the GWs. 

The second scenario corresponds to a magnetic spectrum having a peak at length scale $\lambda_B$, which leads a peak of GWs at the same scale. 

Due  to the effects of inverse cascade, for a broad range of initial conditions, the spectrum evolves to $E_M \propto k^4 $ at small k\cite{Reppin:2017uud,Brandenburg:2017neh}. Thus, in scenario 3, we assume a a Batchelor spectrum $E_M \propto k^4 $ for large scale and a Kolmogorov spectrum $E_M\propto k^{-5/3}$ for small scale. Typically, the helicity $H_M$ can be either positive or negative. For the sake of simplicity, we assume $h=0$. 

For both scenario 2 and scenario 3, the angular spectrum $C_l$ has a peak at $l\sim 2\pi\eta_0/\lambda_B$ or so.

\section{Discussions and conclusion}
\label{sectionconclusion}
In this work, we studied the anisotropies of GWs caused by the inhomogenious weak external hypermagnetic fields. The theory model is the SM+singlet model, where a real scalar field is added as an extension of the SM. Such a new physics model is necessary, in order to offer the strongly first order electroweak phase transition, which can produce stochastic gravitational waves. We have examined the effects of external hypermagnetic fields in the following three steps.

At the first step, We examined the effects of the weak external hypermagnetic field to the bounce solutions and to the electroweak phase transition. It is found that the magnetic field can enhance the value of the 3-D Euclidean action and consequently can decrease the nucleation temperature.

At the second step, we evaluate the GWs spectra in the random magnetic field background. It is found that the GWs can be detected by Taiji, BBO, DECIGO for some parameters of the theory model. Even with the assumption that the external magnetic field is weak, it is found that the hypermagnetic field can further enhance the GWs energy density by about ten percent if we choose $B\sim 800\text{GeV}^2$ at EWPT epoch or $B_0\sim 8\times 10^{-12}\text{G} $ today. The main underlying reason which leads to a significant enhancement of GWs production is due to the decrease of nucleation temperature caused by external hypermagnetic field,  which can enlarge the parameters $\alpha$ and $H_n/\beta$, thus can increase $\Omega$. 


At the third step, we analyzed anisotropies of GWs caused by the magnetic field inhomogeneity. We relate the angular power spectrum of SGWB to the spectrum of magnetic field. Assuming the enhancement of GWs energy density is $\delta_{GW}\sim 0.1$, from the numerical comparison, we observe that the anisotropies can be much larger than the SW and ISW effects come from the primordial density perturbations which are about $l(l+1)C_l^{SW+ISW}/2\pi\sim 10^{-9}$\cite{Li:2021iva}. We also find that for scale invariant magnetic fields, the angular power spectrum $l(l+1)C_l\sim 0.06$ is a constant; for Gaussian-like and random distributed magnetic fields, the angular power spectrum $l(l+1)C_l$ reaches its maximum at  $l\sim 2\pi\eta_0/\lambda_B$ with $l(l+1)C_l<0.06$. In our model, $C_l\propto B^4$, thus it can be secondary if the primordial magnetic fields are extremely weak. For helical hypermagnetic fields, the helicity decay can lead to baryon asymmetry through chiral anomaly. To avoid large amplitude of baryon isocurvature perturbations, magnetic fields at EWPT epoch should less than $1\text{GeV}^2$ as Ref.\cite{Kamada:2020bmb} recommended. If we adopt  the constraint $B<1\text{GeV}^2$, $l(l+1)C_l<1.5\times10^{-13}$, which is negligible.  Our results demonstrate that the detection of the anisotropies of GWs may offer a handle to explore primordial magnetic fields of the universe at the EWPT epoch.

In our study, we have assumed weak external fields in order to find the bounce solutions. It is possible to extend our study to include the case with strong external fields, which can be explored in our future works. 

\appendix

\section{Magnetic corrections to effective potential}
\label{magnetic_correction}
In this work, we have neglected the higher order corrections from magnetic field due to the weak field strength assumption. we will demonstrate  the higher order corrections are negligible in this section. Terms linear to hypermagnetic field in the Lagrangian are
\begin{equation}
	\begin{aligned}
L &\supset -Y_\mu J^\mu\\
J^\mu&=\frac{g_Y}{2}\left(i\partial^\mu H^{\dagger} H-iH^\dagger \partial^\mu H-\sum_iq_i \bar{\psi}_i\gamma^\mu \psi_i\right)		
	\end{aligned}
\end{equation}
To induce a hypermagnetic field, we should add a nontrivial external source $J^\mu$, which means that the plasma in universe are not isotropic. Motivated by the Maxwell equations $J^\mu=\partial_\nu F^{\nu\mu}$, we rewrite the source term as $J^\mu_{ext}=\partial_\nu J_{ext}^{\nu\mu}$, then 
\begin{equation}
	\begin{aligned}
		L \supset-Y_\mu J_{ext}^\mu&=-Y_\mu\partial_\nu J^{\nu\mu}_{ext}\\
		&=-\partial_\nu\left(Y_\mu J_{ext}^{\nu\mu}\right)+\partial_\nu Y_\mu J_{ext}^{\nu\mu}\\
		&=-\partial_\nu\left(Y_\mu J_{ext}^{\nu\mu}\right)+\frac{1}{2}\left(\partial_\nu Y_\mu J^{\nu\mu}_{ext}-\partial_\mu Y_\nu J^{\nu\mu}_{ext}\right)\\
		&=-\partial_\nu\left(Y_\mu J_{ext}^{\nu\mu}\right)+\frac{1}{2}Y_{\nu\mu}J^{\nu\mu}_{ext}. \label{formula2}
	\end{aligned}
\end{equation}
Neglecting the first term which is a surface term, we use the second term to induce hypermagnetic field in Eq. (\ref{source}).
Assuming the hypermagnetic field is along the Z axis, we can simply set $J^{12}_{ext}=-J^{21}_{ext}=B_{ext}$, and other components are zero.

The next order hypermagnetic field contributions come from the one loop corrections for the effective potential. In the symmetric phase, only fermions and Higgs couple to the external field $B_{ext}$.  The propagator for scalar is\cite{Sanchez:2006tt}
\begin{equation}
	\begin{aligned}
		i D_{H}(k) &=\int_{0}^{\infty} \frac{d s}{\cos qB_{ext} s} 
		  \exp \left[i s\left(k_{\|}^{2}-k_{\perp}^{2} \frac{\tan qB_{ext} s}{qB_{ext} s}-m^{2}\right)\right]\\
		  &=\frac{1}{k^{2}-m^{2}}\left(1-\frac{(qB_{ext})^{2}}{\left(k^{2}-m^{2}\right)^{2}}-\frac{2(qB_{ext})^{2} k_{\perp}^{2}}{\left(k^{2}-m^{2}\right)^{3}}+\mathcal{O}(q^2B_{ext}^2)\right).
	\end{aligned}
\end{equation}
The contribution to effective potential from scalars can be computed as 
\begin{equation}
	\begin{aligned}
	V_{H}^1&=\frac{T}{2}\sum_{n,i}\int \frac{d^3k}{(2\pi)^3}\log D^{-1}_H(\omega_n,\vec{k};m_i)\\
	&\simeq \frac{T}{2}\sum_{n,i}\int \frac{d^3k}{(2\pi)^3}\log (\omega_n^2+k^2+m_i^2)\\
	&+(qB_{ext})^2\frac{T}{2}\sum_{n,i}\int \frac{d^3k}{(2\pi)^3}\left[\frac{1}{(\omega_n^2+k^2+m_i^2)^2}-\frac{2k_{\perp}^{2}}{(\omega_n^2+k^2+m_i^2)^3}\right]\\
	&=\frac{T}{2}\sum_{n,i}\int \frac{d^3k}{(2\pi)^3}\log (\omega_n^2+k^2+m_i^2)+(qB_{ext})^2\frac{T}{2}\sum_i\int\frac{1}{3}\frac{d^3k}{(2\pi)^3}\left[\nabla_{\vec{k}}\frac{\vec{k}}{(\omega_n^2+k^2+m_i^2)^2}\right]\,.
	\end{aligned}
\end{equation}
After the renormalization to remove the UV divergence, the first term is what we used in Eq. (\ref{eff_P}). The second term vanishes as demonstrated in Reference \cite{Ayala:2004dx}.
Similarly, the contributions from fermions are  given below\cite{Sanchez:2006tt}
\begin{equation}
	\begin{aligned}
	{ }^{2} V_{f}^{(1)}&=2(qB_{ext})^{2} T \sum_{n} \int \frac{d^{d} k}{(2 \pi)^{d}} \frac{\omega_{n}^{2}+k_{3}^{2}+m_{i}^{2}}{\left(\omega_{n}^{2}+k^{2}+m_{i}^{2}\right)^{3}}\\	
		&=\frac{2(qB_{ext})^{2}}{32 \pi^{2}}\left[\frac{1}{\epsilon}-\frac{7 \xi(3)}{2 \pi^{2}} \frac{m_{i}^{2}}{T^{2}}
		+\ln \left(\frac{4 \mu^{2}}{\pi T^{2}}\right)+\gamma_{E}+\frac{2}{3}\right].
\end{aligned}
\end{equation}
We have used the Mellin summation method\cite{Bedingham:2000ct} to sum the  Matsubara frequencies. After the renormalization procedure, it is found that the contributions of fermions can be put as $f \times (B_{ext})^2$, where the factor $f$ is defined as $f=\frac{q^2}{16 \pi^2}$, which is of size $10^{-2}$ and can be safely omitted when compared with the term given in Eq. (\ref{free_E}).

\begin{acknowledgments}
  We thank Y.D. Chen, A.P. Huang, F.P.Huang, J. D. Shao, and D.W. Wang for helpful discussions. This work is supported in part by the Fundamental Research Funds for the Central Universities, National Natural Science Foundation of China (NSFC) Grant Nos. 12235016, 12221005, 11725523 and 11735007, the Strategic Priority Research Program of Chinese Academy of Sciences under Grant Nos XDB34030000 and XDPB15, the start-up funding from University of Chinese Academy of Sciences (UCAS). The work of Q.S. Yan is supported by NSFC Grant Nos. 11475180 and No. 11875260.
\end{acknowledgments}

\bibliographystyle{unsrt}
\bibliography{myrefs.bib}

\end{document}